# Effects of Synaptic and Myelin Plasticity on Learning in a Network of Kuramoto Phase Oscillators

M. Karimian,[1)] D. Dibenedetto, [1)] M. Moerel, [1,2,3)] T. Burwick,[1,5)] R. L. Westra,[1,4)] P. De Weerd,[1,2,3)] and M. Senden,[2,3)]

[1] *Maastricht Centre for Systems Biology (MaCSBio), Maastricht University, 6229 ER Maastricht, The Netherlands.*

[2]*Department of Cognitive Neuroscience, Faculty of Psychology and Neuroscience, Maastricht University, P.O. Box 616, 6200 MD Maastricht, The Netherlands.*

[3]*Maastricht Brain Imaging Centre, Faculty of Psychology and Neuroscience, Maastricht University, P.O. Box 616, 6200 MD Maastricht, The Netherlands.*

[4] *Department of Data Science and Knowledge Engineering, Maastricht University, 6211 LH Maastricht, The Netherlands.*

[5] *Frankfurt Institute for Advanced Studies, Goethe University Frankfurt, Ruth-Moufang-Str. 1, 60438 Frankfurt am Main, Germany.*

Models of learning typically focus on synaptic plasticity. However, learning is the result of both synaptic and myelin plasticity. Specifically, synaptic changes often co-occur and interact with myelin changes, leading to complex dynamic interactions between these processes. Here, we investigate the implications of these interactions for the coupling behavior of a system of Kuramoto oscillators. To that end, we construct a fully connected, one-dimensional ring network of phase oscillators whose coupling strength (reflecting synaptic strength) as well as conduction velocity (reflecting myelination) are each regulated by a Hebbian learning rule. We evaluate the behavior of the system in terms of structural (pairwise connection strength and conduction velocity) and functional connectivity (local and global synchronization behavior). We find that for conditions in which a system limited to synaptic plasticity develops two distinct clusters both structurally and functionally, additional adaptive myelination allows for functional communication across these structural clusters. Hence, dynamic conduction velocity permits the functional integration of structurally segregated clusters. Our results confirm that network states following learning may be different when myelin plasticity is considered in addition to synaptic plasticity, pointing towards the relevance of integrating both factors in computational models of learning.

**Synaptic and myelin plasticity are two crucial mechanisms underlying learning in the brain. Synaptic plasticity, which refers to activity-dependent changes of synaptic coupling, has been**

**modeled intensely in recent decades. However, myelin plasticity, which refers to activity dependent changes in the structure and thickness of myelin sheaths, has been largely absent from computational models of learning. These two plasticity mechanisms are likely to exhibit complex interactions. In this work we suggest a simple mathematical framework as a first attempt to understand these interactions. Our results may pave the way for the development of new models of learning incorporating both synaptic and myelin plasticity.**

## I. INTRODUCTION

Synchronization, the mutual adjustment of rhythms among interacting oscillators[1,2], is a ubiquitous phenomenon in physics, biology, and neuroscience[3–6]. In the latter, this phenomenon has been linked to various cognitive functions including perception[7–9], attention[10–12], and learning[13–22]. Learning involves the dynamic adjustment of connections among neuronal populations in the form of synaptic plasticity[23]. Mutual interactions between synaptic plasticity and synchronization have been of particular interest in neuroscience[16–22]. However, synaptic plasticity is not the only factor being affected by as well as affecting synchronized activity in oscillating neuronal populations. Myelination is also activity-dependent[24–31] and since it influences the conduction velocity of neuronal signals, it is an additional dynamic factor potentially affecting synchronization behavior. Myelination is integral to the unimpaired functioning of the brain as it ensures that signals originating from presynaptic sources at various locations nevertheless arrive within short succession of each other at a postsynaptic target[32]. The effect of myelination on signal transduction is quite profound with even slight changes in its thickness possessing the ability to bring about significant differences in the number of signals received by a specific neuron within a given time interval[32,33]. This, in turn, might strongly affect local and global synchrony among neural groups. Therefore, it might be beneficial for the brain to dispose of the ability to dynamically adjust signal conduction among remote areas depending on the frequency with which they interact (engage in functional connectivity). Indeed, abundant biological evidence supports the idea of continued adaptive changes in conduction throughout the whole lifespan[26,27,29,34–36]. Given that adaptive myelination constitutes a second dynamic factor in addition to synaptic plasticity, both of which depend on the temporal statistics of neural activations in pre- and post-synaptic neuronal populations[32], we argue that adaptive conduction and synaptic plasticity have reciprocal dependencies and together affect synchronization in the nervous system. These conjoined effects of adaptive myelination and synaptic plasticity on the synchronization behavior of weakly coupled oscillators has so far not been investigated systematically. The present work is intended to fill this gap by supplementing a system of weakly coupled oscillators with both activity-dependent synaptic as well as conduction plasticity and study their impact on synchronization behavior. We employ a neural mass model to capture the phase evolution of weakly coupled neural groups as their connections undergo activity-dependent changes in coupling strength and conduction velocity.

Specifically, we consider a system of Kuramoto oscillators[37] with distance-dependent delays previously established to study the effect of synaptic plasticity[20]. We extend this model by dynamically adjusting conduction velocity (and hence transmission delays) in addition to synaptic weights. Changes in both synaptic weight and conduction depend on a Hebbian learning rule[23], which is based on the frequency of the coactivations among pairs of network oscillators. That is, both connection weights and conduction velocity are time-dependent parameters influencing each other and the dynamics of the network as a whole.

## II. MATERIALS AND METHODS

### A. Weakly-coupled oscillator model

In line with previous work[20], our network model consists of an ensemble of $N$ phase oscillators arranged along a circle; i.e. a one-dimensional array with periodic boundary conditions. The network is fully connected with the exact coupling strengths between oscillators given by the real-valued directed connectivity matrix $K$. Local dynamics of each phase oscillator are governed by a Kuramoto model with transmission delays

$$\dot{\varphi}_i(t) = \omega_i + \frac{1}{N}\sum_{j=1}^{N} K_{ij}(t)\sin\left(\varphi_j(t-\tau_{ij}) - \varphi_i(t)\right) \qquad i = 1, \ldots, N \qquad (1)$$

where $\varphi_i(t) \in [0, 2\pi)$ denotes the phase of oscillator $i$ at time $t$, $\omega_i$ is its intrinsic frequency, $K_{ij}$ reflects the strength of the connection from the $jth$ to the $ith$ oscillator, and $\tau_{ij}$ is the transmission delay from $j$ to $i$ given by $\frac{d_{ij}}{v}$. Here, $d_{ij}$ is the distance between the two oscillators and $v$ is the global conduction velocity. Due to the periodic boundary conditions, the distance can be defined as

$$d_{ij} = \frac{L}{N}\min(|i-j|, N-|i-j|) \qquad (2)$$

with $L$ controlling the circumference of the circle. In accordance with previous work, we define a coupling delay constant $T = \frac{L}{v}$ as the time needed for signals traveling at a velocity $v$ to revolve once around the circle[20]. It follows that

$$\tau_{ij} = \frac{T}{N}\min(|i-j|, N-|i-j|). \tag{3}$$

The coupling strength $K_{ij}$ between oscillators $i$ and $j$ varies dynamically according to a form of Hebbian learning where the growth or decay of coupling strengths depend on the phase offset between oscillators[38,39]

$$\dot{K}_{ij}(t) = \varepsilon_s \left[\alpha_s \cos\left(\varphi_i(t) - \varphi_j(t - \tau_{ij})\right) - K_{ij}(t)\right]. \tag{4}$$

In equation 4, $\varepsilon_s$ and $\alpha_s$ respectively control the learning rate and learning enhancement factor of the coupling strength. The learning enhancement factor $\alpha_s$ determines the maximum and minimum coupling strength[19] and ensures that these remain sufficiently weak.

In addition to scenarios where conduction velocity remains static, we also explore the case in which conduction velocities between pairs of oscillators vary dynamically. That is, conduction velocity is no longer identical for all pairs of oscillators. To that end, we introduce a second Hebbian learning rule accounting for the effects of adaptive myelination

$$\dot{v}_{ij}(t) = \varepsilon_v \left[\alpha_v \cos\left(\varphi_i(t) - \varphi_j(t - \tau_{ij})\right) - v_{ij}(t)\right]. \tag{5}$$

Here, $\varepsilon_v$ and $\alpha_v$ are, respectively, the learning rate and learning enhancement factor of the conduction velocity. Note that conduction velocity was bounded from below at 0.1 because $v_{ij}$ may otherwise grow too small or even become negative if oscillators $i$ and $j$ exhibit an absolute phase offset exceeding $\frac{\pi}{2}$ but $v_{ij} \leq 0$ is not physically meaningful.

### B. Quantitative analyses

#### 1. Global synchronization behavior

In a network of globally coupled oscillators arranged along a ring with distance-dependent delays, the distribution of phases may show propagating structures, referred to as coherent-wave modes[20,40]. That is, phase offsets with respect to a reference oscillator (e.g. the first) may exhibit periodicity at integer (or half-integer, see below) multiples of $2\pi$. Therefore, frequency synchronization in such a system can in principle be characterized in terms of these multiples (denoted by $m$) reflecting the coherent-wave mode of its phase-offsets. However, for the system employed here, identification of coherent-wave mode values is complicated by the fact that it can exhibit synchronization within two anti-phase clusters (double-cluster synchronization) or a single global cluster (single-cluster synchronization). To overcome this problem, we measure phase-coherence assuming a range of candidate modes and select the mode that maximizes phase-coherence. Specifically, phase-coherence is quantified by the order parameters $r_1$ or $r_2$. $r_1$ is defined as

$$r_1 e^{i\psi(t)} = \frac{1}{N} \sum_{j=1}^{N} e^{i[\varphi_j(t) \pm 2\pi m(j-1)/N]} \tag{6}$$

where $\psi(t)$ is the mean phase at time $t$[37]. The order parameter $r_1$, which ranges between 0 and 1, measures the phase-coherence among all oscillators. As such, it can only provide an accurate estimate of phase-coherence in case the system exhibits single-cluster synchronization. In case the system exhibits double-cluster synchronization, a second order parameter, $r_2$ is required[19]

$$r_2{}^2 = |r' - r_1|^2$$

where

$$r' e^{i\psi'(t)} = \frac{1}{N} \sum_{j=1}^{N} e^{2i[\varphi_j(t) \pm 2\pi m(j-1)/N]}. \tag{7}$$

To determine the mode of the system and whether it exhibits single- or double-cluster synchronization in any particular simulation, we compute both $r_1$ and $r_2$ for all candidate mode values ($m \in \{0, 0.5, 1, 1.5, 2\}$) and select the combination with maximum phase-coherence. Please note that for double-cluster synchronization $m$ may take on half-integer values[20].

### *2. Pairwise connectivity*

In addition to the global synchronization behavior of the system, we also examine its local (i.e. pairwise) structural and functional connectivity. Structural connectivity is straightforwardly given by the coupling strength matrix $K$ ranging from $-\alpha_s$ to $\alpha_s$. To measure functional connectivity, we introduce a coherence matrix $D$ whose elements are given by

$$D_{ij} = \frac{1}{\Delta t} \int_{t_r}^{t_r + \Delta t} \cos(\varphi_i(t) - \varphi_j(t)) \, dt. \tag{8}$$

Here, $t_r$ marks a time-point after which the system no longer experiences major changes in coupling strength and/or conduction velocity. $D_{ij}$ ranges from $-1$ to 1 with a value of 1 indicating that two nodes are in phase (over a time interval $\Delta t$) whereas a value of $-1$ indicates that two nodes in anti-phase.

### *3. Numerical simulations*

We analyze the system in terms of its global synchronization behavior as well as in terms of pairwise structural and functional connectivity for three different cases: I) dynamic coupling strength and static conduction velocity; (c.f. [20]) II) static coupling strength and dynamic conduction velocity; and III) dynamic coupling strength and dynamic conduction velocity. For the first scenario, the system is evaluated for a range of combinations of parameters $\varepsilon_s$ and $T$. For the latter two scenarios, $\varepsilon_s$ is fixed at either 0 (no learning, scenario II) or 0.1 (fast learning, scenario III) and the behavior is observed while the parameters $\varepsilon_v$ and $\alpha_v$ are varied. The long-term behavior of the system is characterized by its coherent-wave mode of synchronization and its cluster formation. For notational convenience, we denote each final state $\{m,c\}$, where $m$ indicates the (half-)integer value of the coherent-wave mode and $c$ indicates whether the network exhibits single (s) or double (d) cluster synchronization. For example, state $\{1,d\}$ describes a system exhibiting double cluster synchronization and a mode of 1.

For all simulations, intrinsic frequencies $\omega_i$ are drawn from a normal distribution $\aleph(1,0.01)$ and initial phases are drawn from a uniform distribution in the range $[0,2\pi)$. All simulations start from a network with coupling strengths fixed at their maximum value ($\alpha_s = 1$) which exceeds the critical coupling strength and hence allows for synchronization among oscillators. Furthermore, for those simulations for which velocity changes dynamically, conduction velocities are initialized as $v_{ij}(t = 0) = 0.14$, which means that initial coupling delays correspond to the scenario where the delay constant ($T$) is ~7 for a ring length $L = 1$. Parameters characterizing the network are summarized in table 1 while those characterizing the three simulated scenarios are summarized in table 2.

**Table I:** Network parameters

| Network parameter | value |
|---|---|
| $N$ | 100 |
| $L$ | 1 |

**Table II:** Simulation parameters

| Scenario | parameter | value |
|---|---|---|
| Dynamic coupling strengths, static conduction velocities | $\alpha_s$ | 1 |
| | $\varepsilon_v$ | 0 |
| | $\alpha_v$ | 0 |
| Static coupling strengths, dynamic conduction velocities | $\varepsilon_s$ | 0 |
| | $\alpha_s$ | 1 |
| Dynamic coupling strengths and conduction velocities | $\varepsilon_s$ | 0.1 |
| | $\alpha_s$ | 1 |

The model is implemented in MATLAB (R2016a) and integrated for 20000 time steps using the forward Euler method with a step size $dt = 0.01$ in arbitrary units of time. To accommodate for delays, we always first simulate 1000 time steps during which oscillators are non-interacting. Subsequently, the time delay interaction is switched on to simulate the 19000 time steps of interest.

We perform 50 simulations with different randomizations of initial conditions for each parameter combination in every scenario. We select the most frequently observed combination of coherent-wave mode of synchronization and cluster-formation (single vs double) as the characteristic final state of a given parameter combination. Whenever the characteristic state is observed in less than 70% of the simulations, we additionally identify a secondary state as the one occurring for at least 50% of the remaining simulations (i.e., of those not classified as the characteristic state). If no secondary state can be unambiguously identified, we regard the state as uncharacterizable. This procedure assumes that states are discernible for individual simulations; that is, they are indeed characterizable in terms of coherent-wave mode of synchronization and cluster-formation. If this assumption is violated, we regard the system as erratic.

## III. RESULTS

### A. Scenario I: dynamic coupling strengths, static conduction velocities

We first examined learning in the context of static conduction velocity. For this purpose, we explored a parameter space defined by the delay constant $T$ and the learning rate $\varepsilon_s$. Most parameter settings yielded highly consistent results. However, some regions of parameter space exhibit diverse results. This is especially prevalent near borders between adjacent regions and likely reflects transitions in mode synchronization, cluster-formation, or both. The two parameters affect the behavior of the system in different, albeit interacting, ways. The learning rate mainly affects cluster-formation, with slow learning leading to the emergence of a single cluster while fast learning leads to the formation of two clusters (see figure 1a). In the former case, changes in coupling strength between pairs of oscillators occur at a slower rate than synchronization. That is, the system synchronizes before large initial phase offsets can decrease coupling. In the latter case, changes in coupling strength between pairs of oscillators occur at a faster rate than synchronization. That is, initially large phase offsets between pairs of oscillators quickly drive their coupling strength to negative values, thus exacerbating their offset until they are separated by exactly $\pi$.

The delay constant, on the other hand, mainly affects mode synchronization with longer delays leading to larger $m$ (see figure 1). Specifically, for non-zero values, phases distribute around the circle such that
77

the offset between each pair of neighboring oscillators is $\frac{2\pi}{N}m$ (within a cluster) or $\frac{2\pi}{N}m + \pi$ (across clusters). Note that for the emergence of two clusters, half-integer values can be obtained (figure 1d,f). This is in line with previous observations[20] that half-integer values are the result of the two clusters interconnecting. Oscillator pairs within a cluster "*see*" each other in phase when their phase offsets are matched by their delays. That is, due to delays, from the perspective of each oscillator in a cluster, the other oscillators within the same cluster appear in-phase whereas to an external observer they may appear out of phase. For the emergence of a single cluster, there is an exception to this observation for oscillator pairs with a phase offset around $\frac{\pi}{2}$. For these values, the trailing oscillator sees the leading oscillator in phase. However, the leading oscillator sees the trailing one in anti-phase. This asymmetry affects the coupling strength such that the structural connection from the leading to the trailing oscillator is positive while that from the trailing to the leading is negative. The magnitude of their coupling strength is otherwise equal. This leads to one or two stripes of negative values in the structural connectivity matrix for modes $m = 1$ and $m = 2$, respectively (see figure 2g,h). Interestingly, the structural connectivity matrices emerging for double-cluster formation also exhibit stripes for non-zero modes (figure 2d-f). The number of these stripes in each case is twice its corresponding mode value $m$. According to the Hebbian learning rule (equation 4), coupling strengths between every two oscillators $i$ and $j$ approach a stable value given by $K_{ij} = \alpha_s \cos(\varphi_i - \varphi_j)$. For phase differences of $(2n - 1)\frac{\pi}{2}$ this entails that the connection weights between the corresponding oscillators decay to zero. Since the mode determines the repetition of phase offsets equal to $(2n - 1)\frac{\pi}{2}$ for each oscillator, it also determines the number of stripes in the structural connectivity matrices.

The emergence of stripes is also apparent in functional connectivity matrices (figure 3). Here, stripes are symmetric, however, since functional connectivity is undirected. Therefore, twice as many stripes can be observed in functional connectivity matrices as compared to structural connectivity matrices. Furthermore, the exact location of stripes in the structural and functional connectivity matrices are different because temporal delays are not considered in the computation of pairwise correlations.

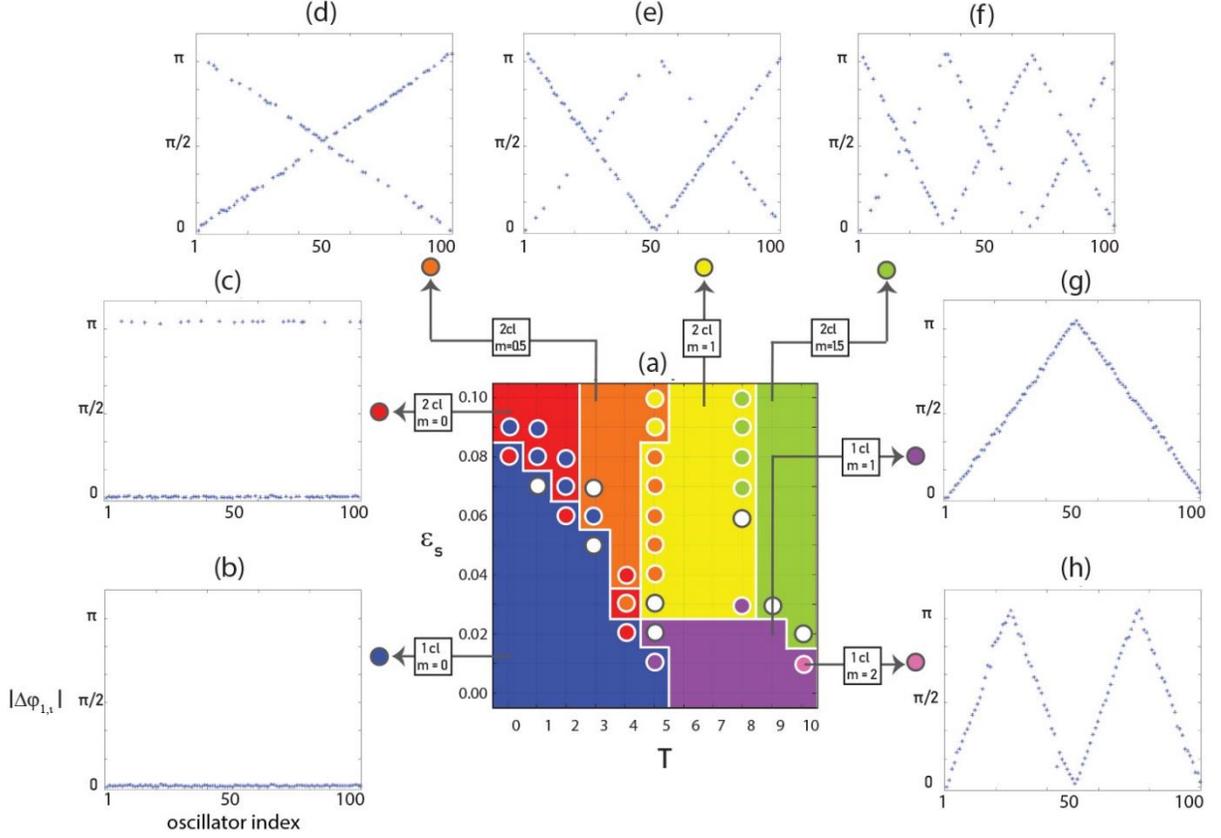

**Figure 1: Arrangement of phase offsets with respect to the first oscillator when coupling strength is dynamic and conduction velocity is static. Panel a)** shows the color-coded state (coherent-wave mode of synchronization and cluster-formation) for each point in the parameter space defined by $T$ and $\varepsilon_s$. Colors indicate the characteristic states. Furthermore, colored disks indicate secondary states. A white disk indicates that the state was uncharacterizable. **Panel b)** shows absolute phase offsets between every oscillator and the first $\left(\left|\Delta\varphi_{1,i}\right|\right)$ for state{0,s}. All offsets are close to zero. **Panel c)** shows $\left|\Delta\varphi_{1,i}\right|$ for the state{0,d}. Phase offsets are close to zero for oscillators falling into the same cluster as the first and close to $\pi$ (half period) for those falling into the opposite cluster. **Panel d)** shows $\left|\Delta\varphi_{1,i}\right|$ for the state{0.5,d}. Phase offsets exhibit one half-cycle; i.e. oscillators falling into the same cluster as the first increases with distance, whereas those in the opposite cluster decrease with distance. **Panel e)** shows $\left|\Delta\varphi_{1,i}\right|$ for the state{1,d}. Phase offsets exhibit one full cycle with offsets for oscillators falling into the same cluster as the first mirroring those of oscillators in the opposite cluster. **Panel f)** shows $\left|\Delta\varphi_{1,i}\right|$ for the state{1.5,d}. Phase offsets exhibit one and a half cycles with offsets for oscillators falling into the same cluster as the first mirroring those of oscillators in the opposite cluster. **Panel g)** shows $\left|\Delta\varphi_{1,i}\right|$ for the state{1,s}. Phase offsets exhibit one full cycle. **Panel h)** shows $\left|\Delta\varphi_{1,i}\right|$ for the state{2,s}. Phase offsets exhibit two full cycles passed by a single cluster. All phase offsets are averaged over the last 100 time steps. Phase offsets for each parameter combination are shown in supplementary figure S1b.

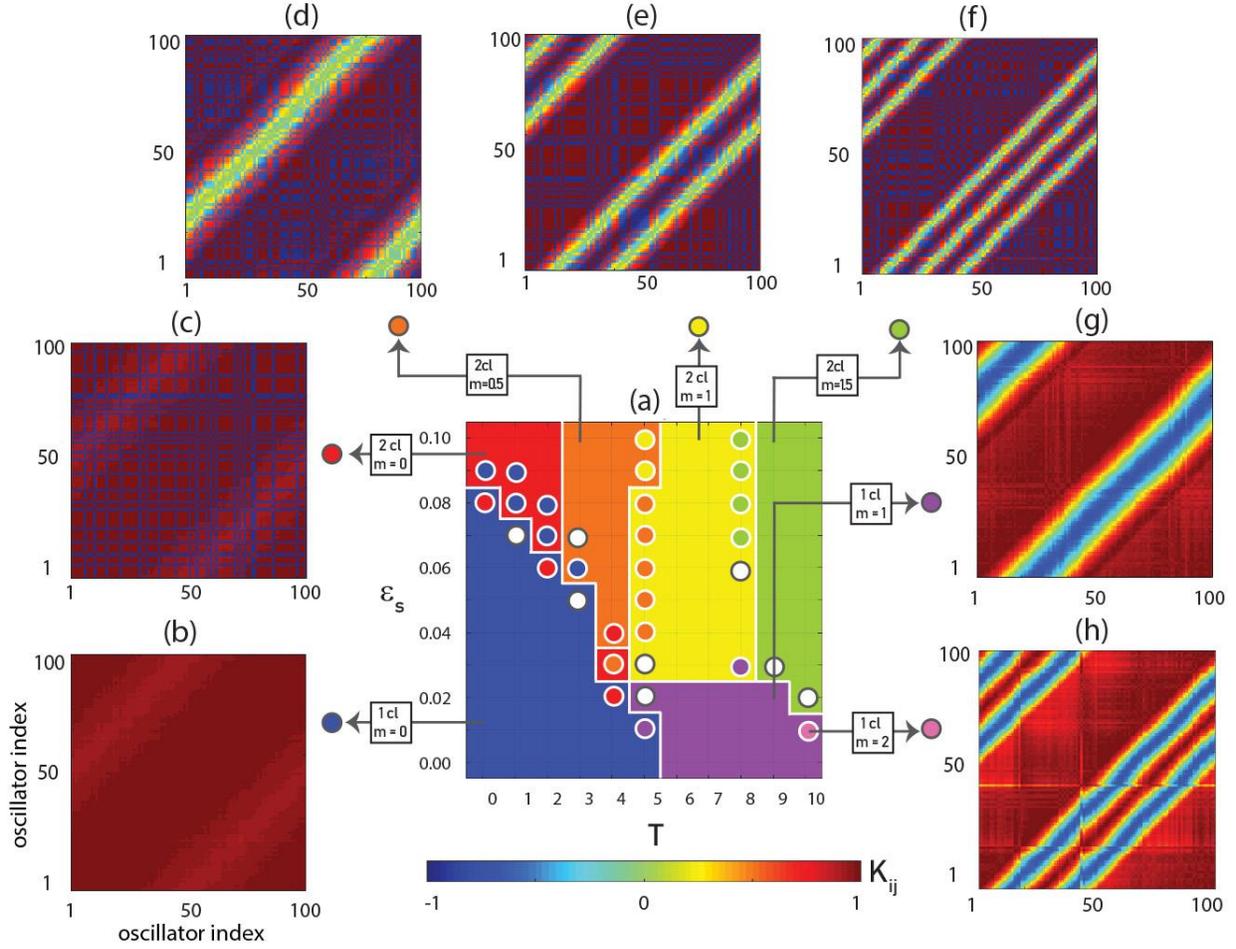

**Figure 2: Pairwise structural connectivity emerging in the context of dynamic coupling and static conduction. Panel a)** shows the color-coded state of coherent-wave mode of synchronization and cluster-formation observed at each point in the parameter space defined by $T$ and $\varepsilon_s$. As in figure 1, the secondary state is marked with colored disks where white indicates uncharacterizable states. **Panel b)** shows the structural connectivity matrix of the network for the state{*0,s*}. The network largely preserves the initial connectivity pattern. **Panel c)** shows structural connectivity of the network for the state{*0,d*}. Pairwise connection weights are close to $+\alpha_s$ and $-\alpha_s$ for oscillator pairs belonging to the same or distinct clusters, respectively. **Panels d-f)** show structural connectivity matrices of the network for the state{*0.5,d*} (panel d), state{*1,d*} (panel e), state{*1.5,d*} (panel f). As before, coupling weights have approached $+\alpha_s$ for within cluster connections and $-\alpha_s$ for between cluster connections. However, based on the mode synchronization, 1, 2 and 3 stripes of near-zero connection weights have formed in panels d, e and f, respectively. **Panel g)** shows the structural connectivity matrix of the network for the state{*1,s*}. All possible phase offsets $((n-1)(2\pi/N))$ with respect to the first oscillator can be observed. **Panel h)** shows the structural connectivity matrix for a network given the state{*2,s*}. The same observations as for panel g can be made, with the difference that phase differences are repeated. The structural connectivity matrices are averaged over the last 100 time steps of the simulation. Structural connectivity matrices for each parameter combination are shown in supplementary figure S1c.

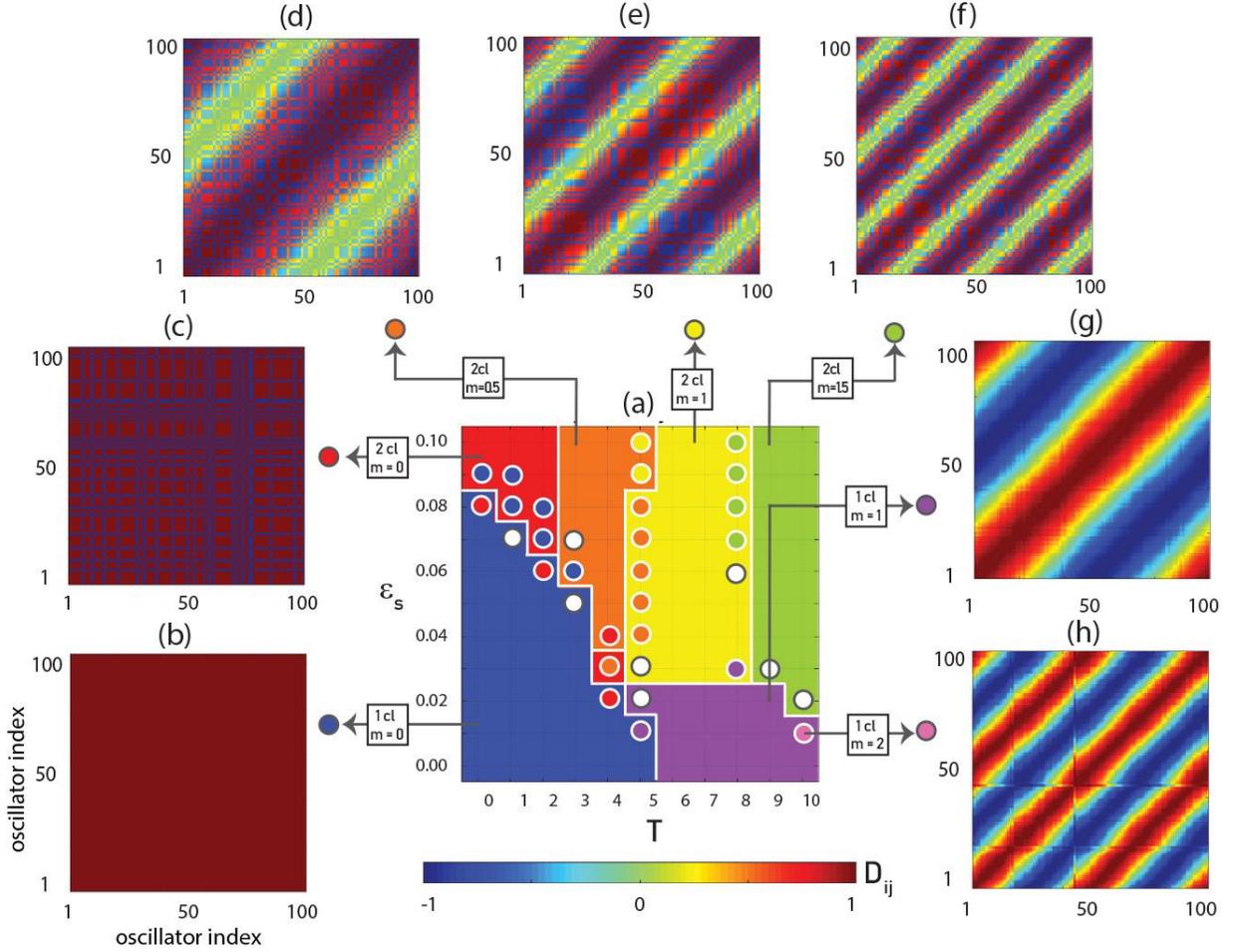

**Figure 3: Pairwise functional connectivity among oscillators emerging when coupling strength is dynamic and conduction is static**. **Panel a)** shows the color-coded state of coherent-wave mode of synchronization and cluster-formation observed at each point in the parameter space defined by $T$ and $\varepsilon_s$. Color coding is the same as in figure 1. **Panel b)** shows the functional connectivity matrix of the network for the state$\{0,s\}$. The globally correlated functional connectivity matrix resembles the structural connectivity matrix. **Panel c)** shows the functional connectivity matrix of a network for the state$\{0,d\}$. **Panels d-f)** show functional connectivity matrices of networks for the state$\{1.5,d\}$ (panel d), state$\{1,d\}$ (panel e), state$\{1.5,d\}$ (panel f). The functional pairwise correlations are associated with the cluster-formation of oscillators as they are 1 or close to 1 for intra-cluster correlations and are $-1$ or close to $-1$ for between cluster correlations. Based on the mode of synchronization, 2, 4 and 6 stripes of zero or very weak correlations in panel d, e and f are formed, respectively. **Panel g)** shows the functional connectivity matrix of a network for the state$\{1,s\}$. Pairwise functional connectivity values are 1 for the neighboring oscillators and decrease to $-1$ as a function of distance. **Panel h)** shows the functional connectivity matrix of the network for the state$\{2,s\}$. A similar pattern as for panel g manifests, but reflecting two complete revolutions of phase offsets around the circle. The elements of correlation matrices were computed over the last 100 time steps of the simulation. Functional connectivity matrices for each parameter combination are shown in supplementary figure S1d.

## B. Scenario II: static coupling strengths, dynamic conduction velocities

Next, we examine the effects of dynamic conduction velocity on a network with static connection weights to establish the unique effects of adaptive myelination on functional connectivity among phase oscillators. To that end, we vary the learning rate $\varepsilon_v$ and enhancement factor $\alpha_v$ controlling dynamic changes in conduction velocity. Note that we no longer vary the coupling delay constant $T$ since delays depend on conduction. Rather, we initialize conduction velocity among oscillator pairs such that $v_{ij}(t=0) = 0.14$, which means that the initial coupling delays correspond to the case where $T \cong 7$. These parameter settings correspond to a system exhibiting state{$1,s$} in simulations where conduction remains static. For dynamic conduction velocity, state{$1,s$} is still observed most frequently irrespective of which values have been chosen for $\varepsilon_v$ and $\alpha_v$. However, within a contiguous region of parameter space, the system exhibits state{$2,d$} as its secondary state (figure 4a). Furthermore, at the borders of this region, the system exhibits highly variable behavior rendering its state uncharacterizable.

Figure 4 shows absolute phase offsets of all oscillators with respect to the first. Remarkably, for state{$2,d$} phases cluster around 0 and $\pi$ with sharp transitions between the two rather than smooth transitions. In fact, dynamic conduction velocity pushes phase offsets to either 0 or $\pi$ which brings about a transformation from state{$2,s$} to state{$2,d$}. This localized clustering leads to highly structured clusters, where an oscillator's affiliation with a cluster is determined by its location along the ring. Interestingly, conduction matrices emerging for state{$2,d$} suggest that the system exhibits four distinct clusters rather than two (see figure 5d); one cluster for each peak and trough of the phase offsets (cf. figure 4d). That is, signals are conducted fast among oscillators within a peak (trough) and slow among oscillators across peaks (troughs). This is the result of initial conditions. With conduction velocity being equal, short distances among oscillators within a peak (trough) lead to short delays, whereas long distances across peaks (troughs) lead to long delays. In this case, the pressure to synchronize peaks (troughs) is most easily met when signals are transmitted instantaneously within a peak (trough) or with a delay matching exactly one period across peaks (troughs). Functionally, these four clusters are not discernible (see figure 6d) since oscillators falling into both peaks (troughs) exhibit no phase-offset with respect to each other.

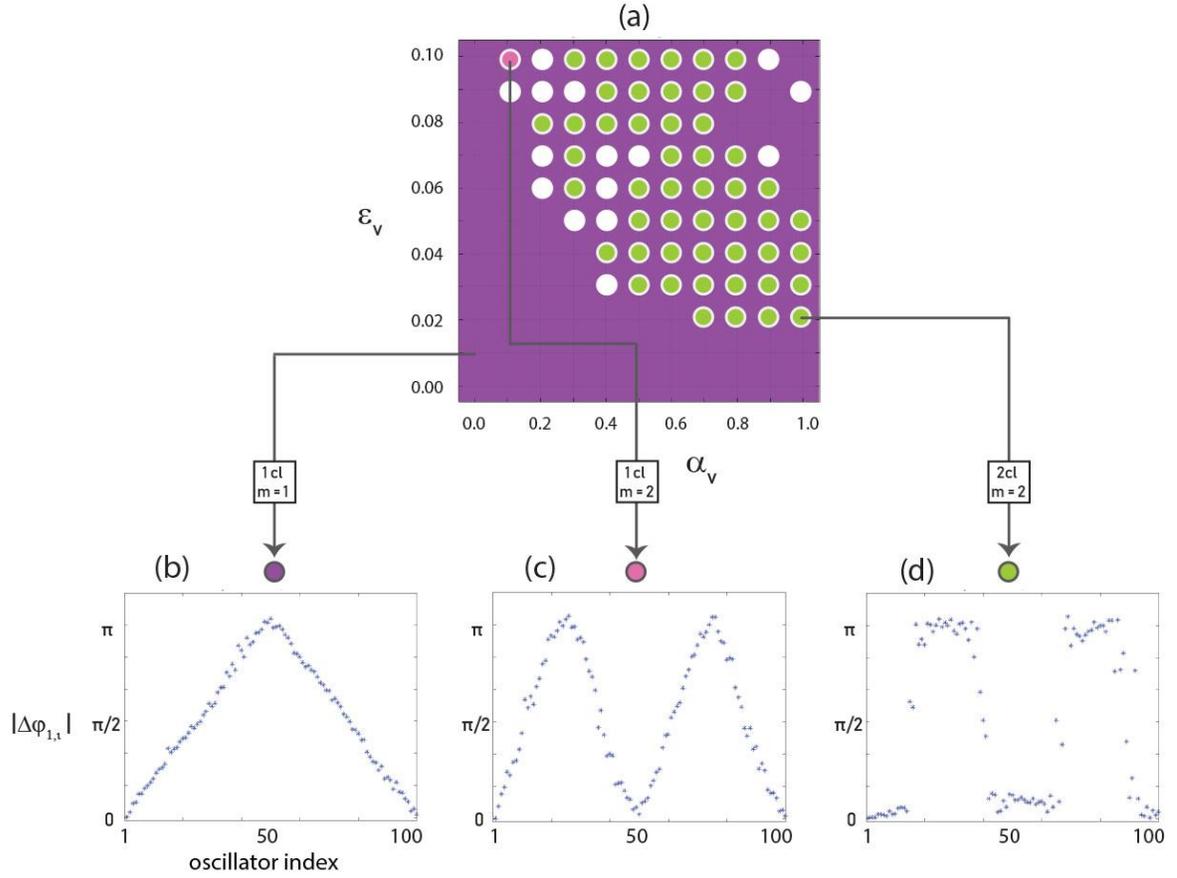

**Figure 4: Phase offsets with respect to the first oscillator when coupling strength is static and conduction is dynamic. Panel a)** shows the color-coded state of coherent-wave mode of synchronization and cluster-formation observed at each point in the parameter space defined by $\varepsilon_v$ and $\alpha_v$. Color coding is the same as in figure 1. The entire parameter space is primarily characterized by state$\{1,s\}$. However, a wide region of parameter space exhibits a secondary state defined by a mode of 2 and the formation of two clusters. **Panel b)** shows $|\Delta\varphi_{1,i}|$ for state$\{1,s\}$. Phase offsets exhibit one full cycle. **Panel c)** shows $|\Delta\varphi_{1,i}|$ for state$\{2,s\}$. Phase offsets exhibit two full cycles. **Panel d)** shows $|\Delta\varphi_{1,i}|$ for state$\{2,d\}$. Phase offsets are largely pushed to either 0 or $\pi$, depending on cluster affiliation. All phase offsets are averaged over the last 100 time steps. Phase offsets for each parameter combination are shown in supplementary figure S2b.

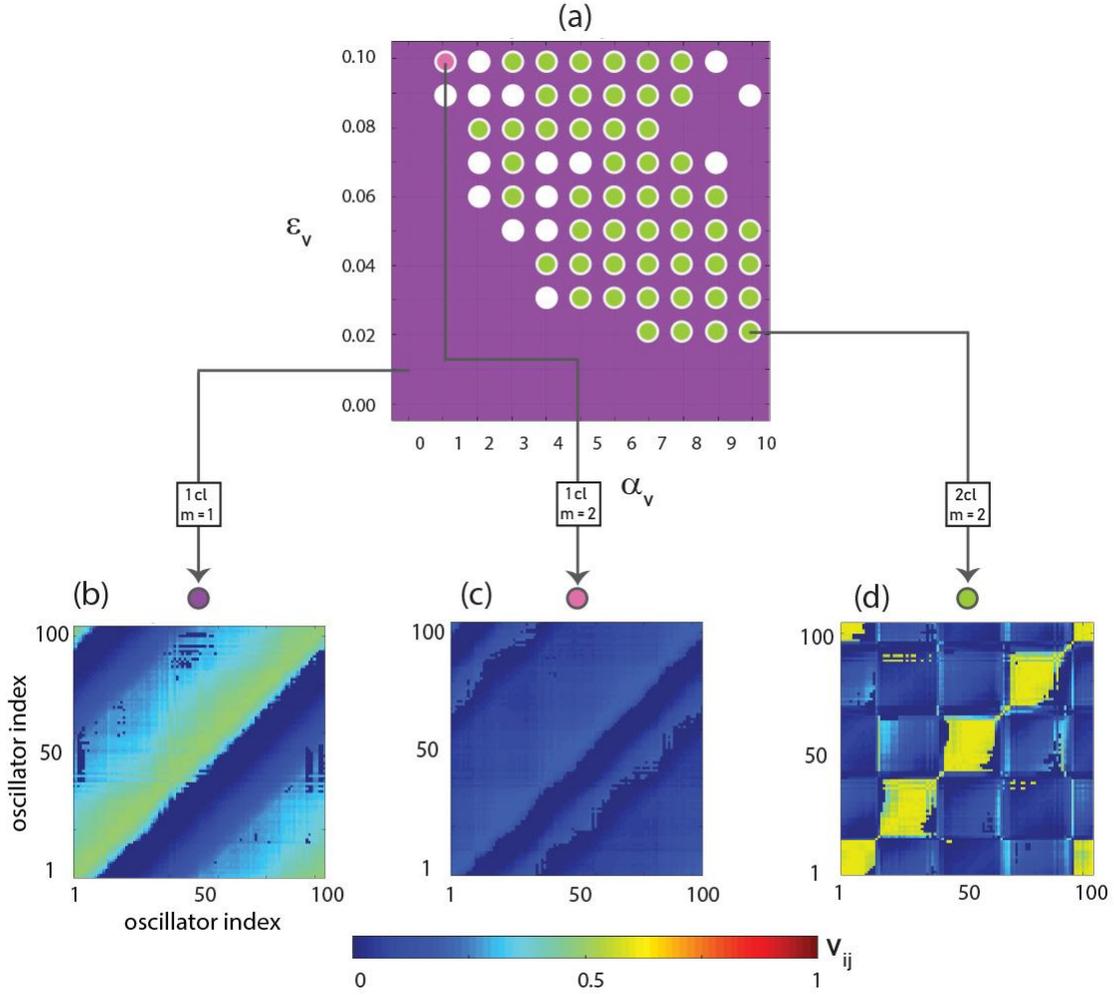

**Figure 5: Conduction velocity matrices when coupling strength is static and conduction is dynamic. Panel a)** shows the color-coded state of coherent-wave mode of synchronization and cluster-formation observed at each point in the parameter space defined by $\varepsilon_v$ and $\alpha_v$. Color coding is the same as in figure 1. **Panels b and c)** show the pairwise conduction velocity matrices for state$\{1,s\}$ (reflecting one full cycle of phase offsets), and state$\{2,s\}$ (reflecting two full cycles of phase offsets), respectively. **Panel d)** shows the pairwise conduction velocity matrices for state$\{2,d\}$. Conduction velocities between the intra--cluster oscillators are noticeably higher than those between other pairs. The conduction velocity matrices are averaged over the last 100 time steps of the simulation. Conduction velocity matrices for each parameter combination are shown in supplementary figure S2c.

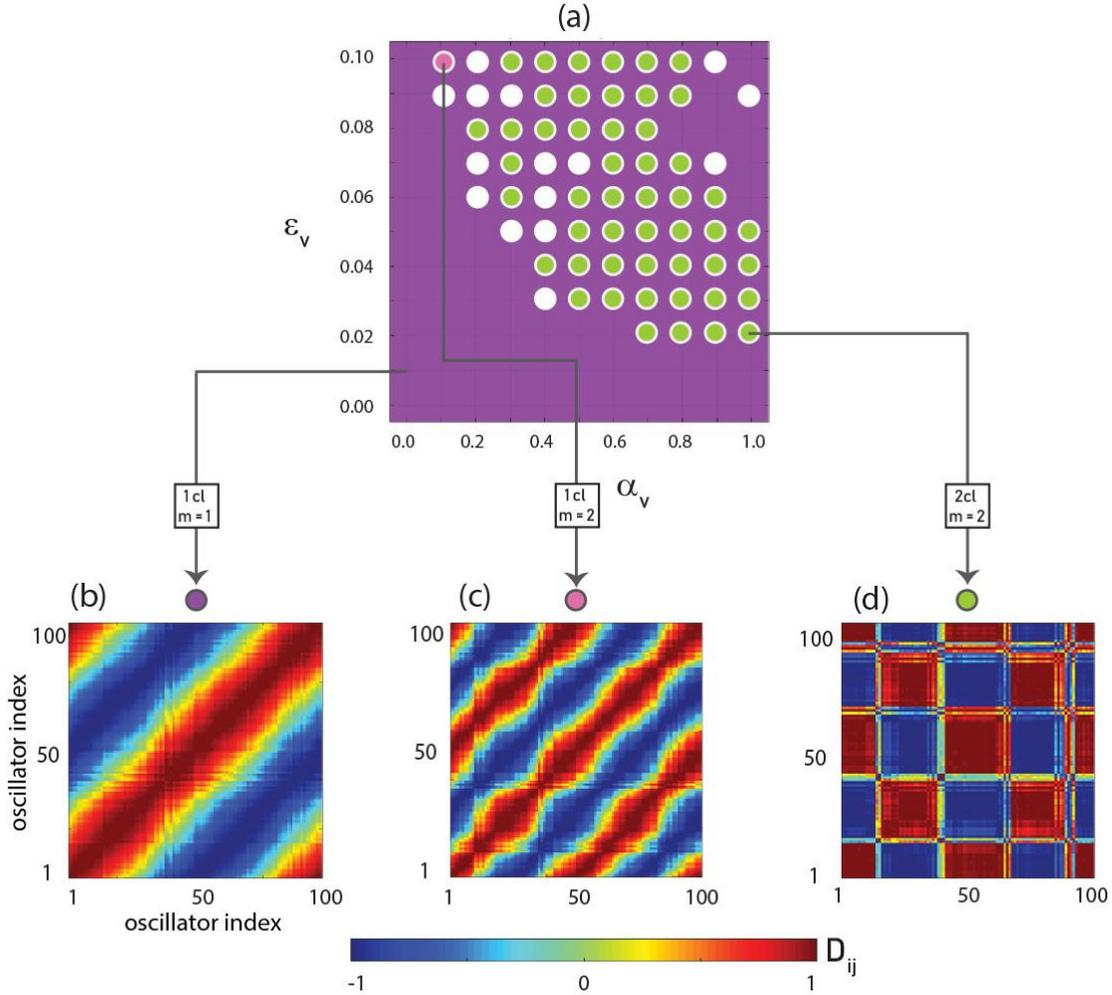

**Figure 6: Pairwise functional connectivity among oscillators when coupling strength is static and conduction is dynamic.** **Panel a)** shows the color-coded state of coherent-wave mode of synchronization and cluster-formation observed at each point in the parameter space defined by $\varepsilon_v$ and $\alpha_v$. Color coding is the same as in figure 1. **Panel b)** shows a representative functional connectivity matrix of the network for state{$1,s$}. The matrix reflects a full cycle of phase offsets. **Panel c)** shows a functional connectivity matrix of the network for state{$1,s$}. Two complete revolutions of the relative phase offsets are exhibited. **Panel d)** shows a functional connectivity matrix of the network for state{$2,d$}. A vast majority of the pairwise correlations reflect either in-phase or anti-phase relations among oscillators. The correlation matrices were computed over the last 100 time steps of the simulation. Functional connectivity matrices for each parameter combination are shown in supplementary figure S2d.

### C. Scenario III: dynamic coupling strengths and conduction velocities

Having explored the effects of dynamic structural connectivity and dynamic conduction velocity in isolation, we next investigate their interaction. Dynamic changes in connection strength and conduction velocity constitute the most biologically relevant scenario. In this simulation, initial values of the conduction velocity matrix $v$ were again chosen such they resemble the condition where $T \cong 7$. Furthermore, the learning rate $\varepsilon_s$ was fixed at 0.1 (fast learning). Recall that this configuration produces

state{$1,d$} for static conduction velocity (cf. figure 1a). As for scenario II, we explore the parameter space defined by the enhancement factor $\alpha_v$ and the learning rate $\varepsilon_v$ controlling dynamic conduction velocity. Figure 7a reveals that the behavior of the system is mainly affected by the enhancement factor $\alpha_v$. If the learning rate $\varepsilon_v$ is small, conduction velocity changes too slowly to have any discernible influence on the behavior of the system and state{$1,d$} is preserved for all values of $\alpha_v$. Once the conduction learning rate $\varepsilon_v$ is sufficiently large (though it may still be a factor of 10 smaller than the learning rate controlling synaptic coupling strength), however, the behavior of the system is entirely determined by $\alpha_v$.

For values of $\alpha_v < 0.14$, conduction necessarily decays towards values lower than initialization. This produces a situation essentially equivalent to fast learning and very long delays ($T \geq 9$) in scenario I and leads to the emergence of state{$1.5,d$} (cf. figure 1f). For $\alpha_v \cong 0.14$, the system exhibits erratic behavior. To account for the system's behavior as $\alpha_v$ increases, it is essential to consider the fact that both coupling strengths and conduction velocities evolve according to the same Hebbian learning rule with the sole difference that conduction velocities are bounded from below at 0.1. This implies that whenever the coupling strength between two oscillators tends towards $+\alpha_s$, conduction velocity between the two increases (towards $\alpha_v$). In contrast, whenever the coupling strength between two oscillators tends towards $-\alpha_s$, coupling velocity between the two decreases (towards 0.1). This implies that coupling strength and conduction velocity act agonistically for oscillators within the same cluster; these oscillators are both positively coupled and exhibit fast conduction velocity (short delays). However, for oscillators in separate clusters, coupling strength and conduction velocity act antagonistically. Negative coupling is paired with slow conduction velocity (long delays). For intermediate values of $\alpha_v$, oscillators in different clusters see each other in anti-phase for phase offsets smaller than $\pi$. They thus form two clusters whose offset is less than half a period (figure 7e). For large values of $\alpha_v$, oscillators in different clusters see each other in anti-phase for phase offsets close to zero (figure 7d). This allows them to form a single functional cluster (figure 10d) even though structurally, both in terms of coupling strength (figure 8d) and conduction velocity (figure 9d), they form separate clusters. The system can thus exhibit a wide array of states not observed when considering dynamic coupling strength alone.

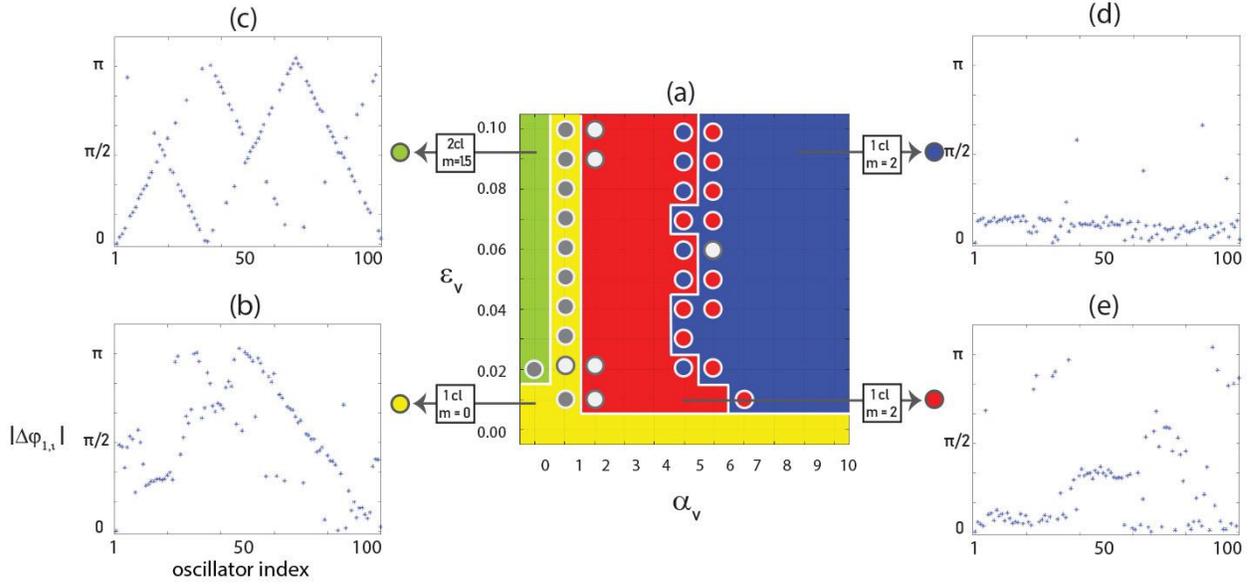

**Figure 7: Phase offsets with respect to the first oscillator when coupling strength and conduction velocity are both dynamic. Panel a)** shows the color-coded state of coherent-wave mode of synchronization and cluster-formation observed at each point in the parameter space defined by $\varepsilon_v$ and $\alpha_v$. Color coding is the same as in figure 1. Gray circles mark erratic states. **Panel b)** shows $|\Delta\varphi_{1,i}|$ for state{1,d}. **Panel c)** shows $|\Delta\varphi_{1,i}|$ for state{1.5,d}. Phase offsets exhibit one and a half cycles **Panel d)** shows $|\Delta\varphi_{1,i}|$ for state{0,s}. Aside from a few exceptions, offsets are generally close to zero. **Panel e)** shows $|\Delta\varphi_{1,i}|$ for state{0,d}. While our procedure identified this example as 0-mode synchronization, visually it appears to not fit any state particularly well. Phase offsets were averaged over the last 100 time steps. Phase offsets for each parameter combination are shown in supplementary figure S3b.

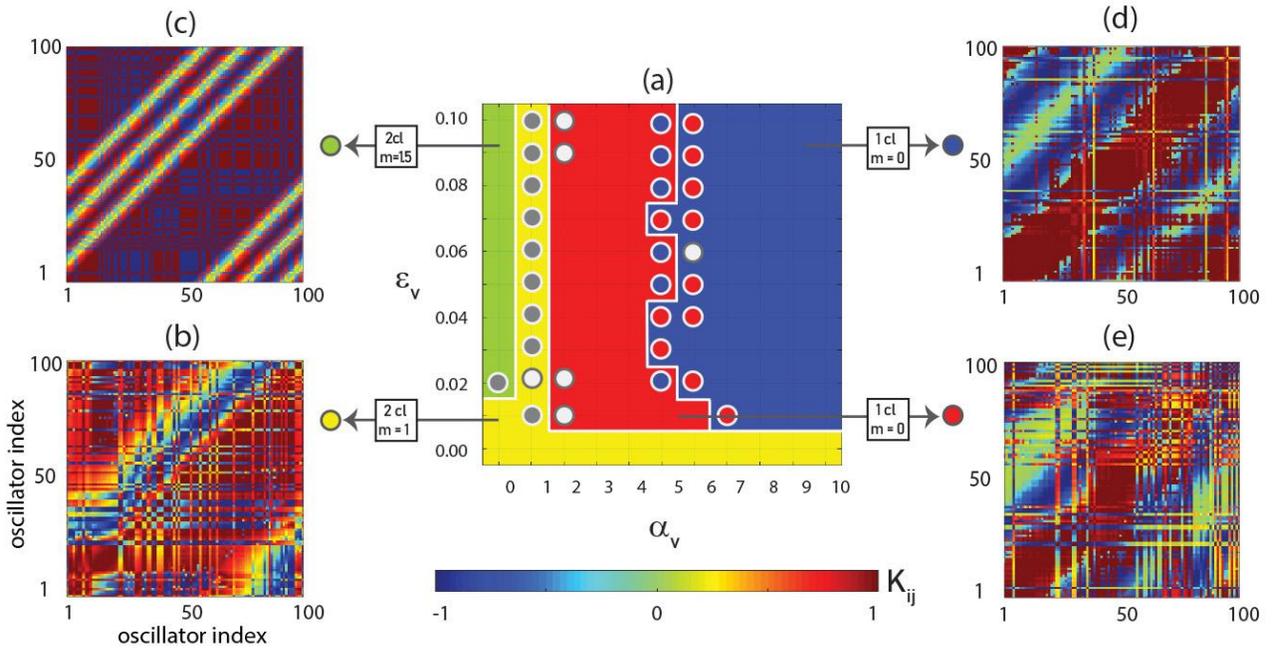

**Figure 8: Pairwise structural connectivity emerging when coupling strength and conduction velocity are both dynamic. Panel a)** shows the color-coded state of coherent-wave mode of synchronization and cluster-formation observed at each point

in the parameter space defined by $\varepsilon_v$ and $\alpha_v$. Color coding is the same as in figure 1 (gray disks as in figure 7). **Panels b)** shows structural connectivity of the network for state{*1,d*}. **Panel c)** shows structural connectivity matrix of the network for state{*1.5,d*}. As for simulations with static conduction velocity, in this region, connectivity matrices exhibit 3 (2*m*) stripes reflecting weak connections. **Panel d)** shows structural connectivity of the network for state{*0,s*}. **Panel e)** shows structural connectivity of the network for state{*0,d*}. The structural connectivity matrices are averaged over the last 100 time steps of the simulation. Structural connectivity matrices for each parameter combination are shown in supplementary figure S3c.

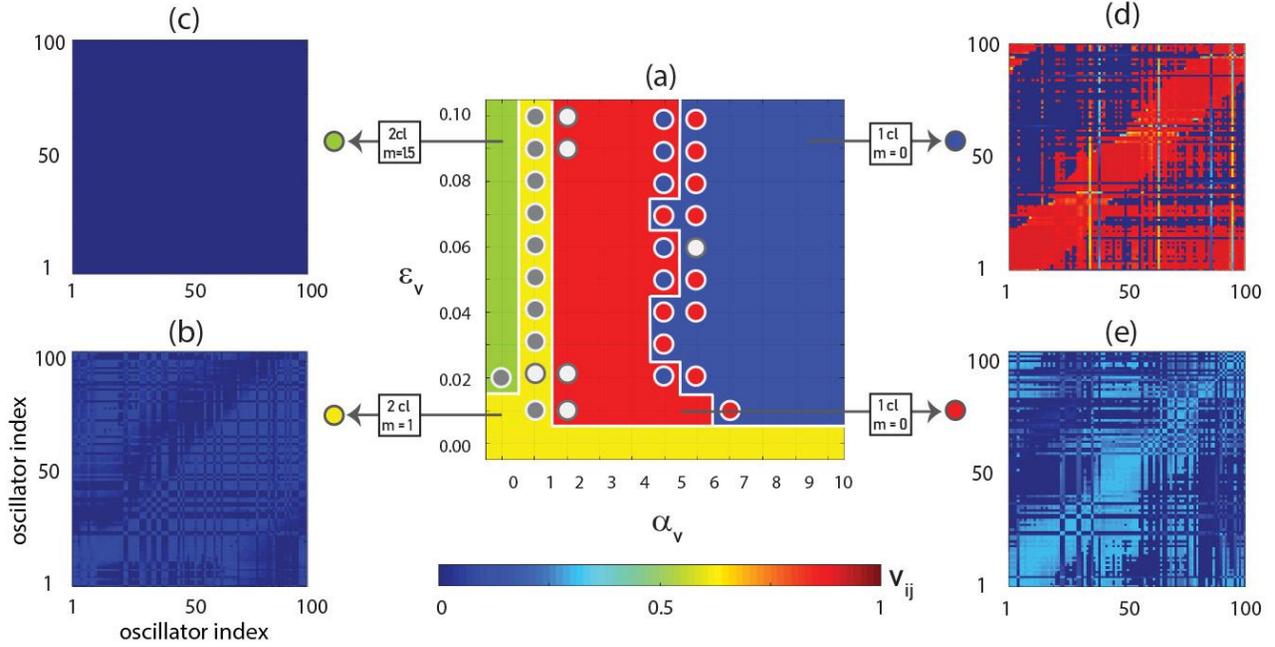

**Figure 9: Pairwise conduction velocities among oscillators when coupling strength and conduction velocity are both dynamic. Panel a)** shows the color-coded state of coherent-wave mode of synchronization and cluster-formation observed at each point in the parameter space defined by $\varepsilon_v$ and $\alpha_v$. Color coding is the same as in figure 1 (gray disks as in figure 7). **Panel b)** shows the pairwise conduction velocity of the network for state{*1,d*}. Conduction velocities only change slightly relative to their initial values. **Panel c)** shows pairwise conduction velocity of the network for state{*1.5,d*}. Conduction velocities have decayed to zero. **Panel d)** shows pairwise conduction velocity of the network for state{*0,s*}. **Panel e)** shows pairwise conduction velocity of the network for state{*0,d*}. The conduction velocity matrices are averaged over the last 100 time steps of the simulation. Conduction velocity matrices for each parameter combination are shown in supplementary figure S3d.

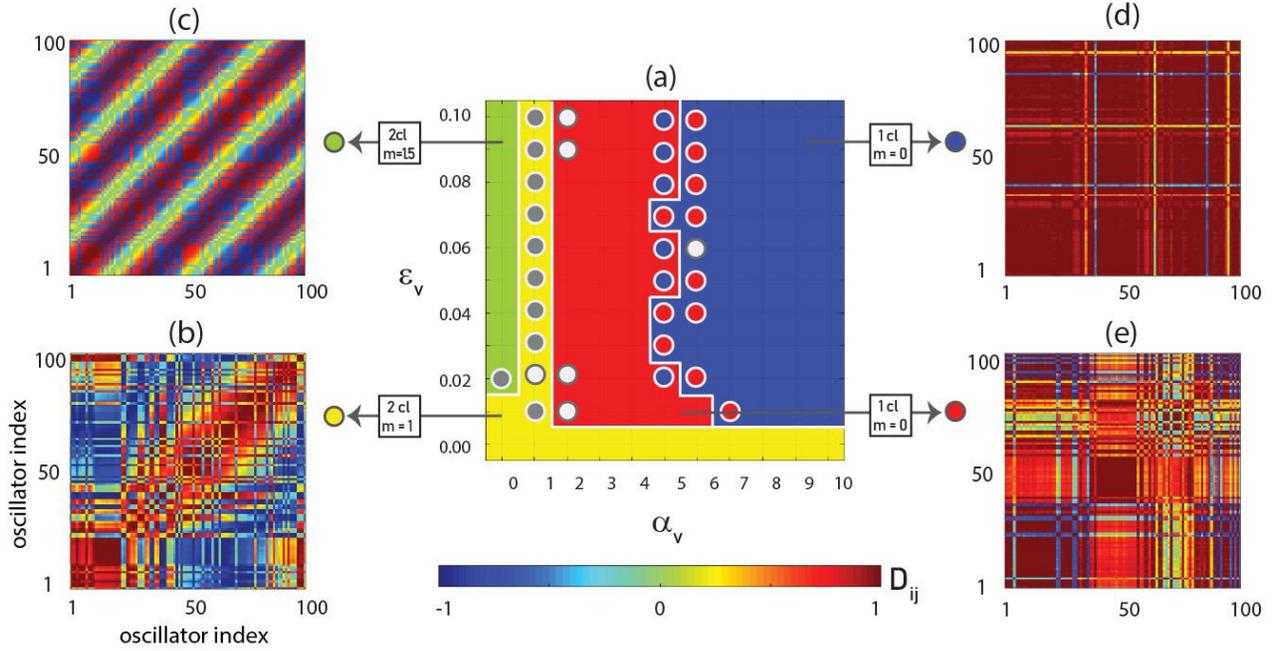

**Figure 10: Pairwise functional connectivity among oscillators when coupling strength and conduction velocity are both dynamic. Panel a)** shows the color-coded state of coherent-wave mode of synchronization and cluster-formation observed at each point in the parameter space defined by $\varepsilon_v$ and $\alpha_v$. Color coding is as in figure 1 (gray disks as in figure 7). **Panel b)** shows functional connectivity of the network for state$\{1,d\}$. **Panel c)** shows functional connectivity of the network for state$\{1.5,d\}$. The formation of $4m$ stripes of zero or very weak connection weights can be observed. **Panel d)** shows the functional connectivity matrix for a network of state$\{0,s\}$. **Panel e)** shows the functional connectivity matrix for a network of state$\{0,d\}$. Correlation matrix elements are averaged over the last 100 time steps of the simulation. Functional connectivity matrices for each parameter combination are shown in supplementary figure S3e.

## IV. DISCUSSION

In the present study we investigated the effects of synaptic plasticity (dynamic coupling strength) and adaptive myelination (dynamic conduction velocity) on the synchronization behavior of weakly coupled oscillators arranged on a circle. For dynamic coupling strength combined with static conduction velocity, we found that depending on the learning rate controlling changes in coupling strength, a single or two clusters can emerge. Furthermore, depending on delay, phase offsets may exhibit periodicity according to coherent-wave modes of synchronization. For non-zero modes, structural clusters become functionally apparent only after correcting for offsets. For zero modes, a tight correspondence between structural and functional clusters is straightforwardly apparent. This is no longer the case once conduction velocity is allowed to vary. Already in the context of static coupling strength, we observed that dynamic conduction velocity dissociates structural from functional connectivity as a wide range of parameter combinations allow for the emergence of four structural clusters (in terms of pairwise synaptic conduction velocity) which functionally present as two clusters. That is, if conduction velocity is dynamic, clusters may be structurally distinct while functionally connected (see figure 11).

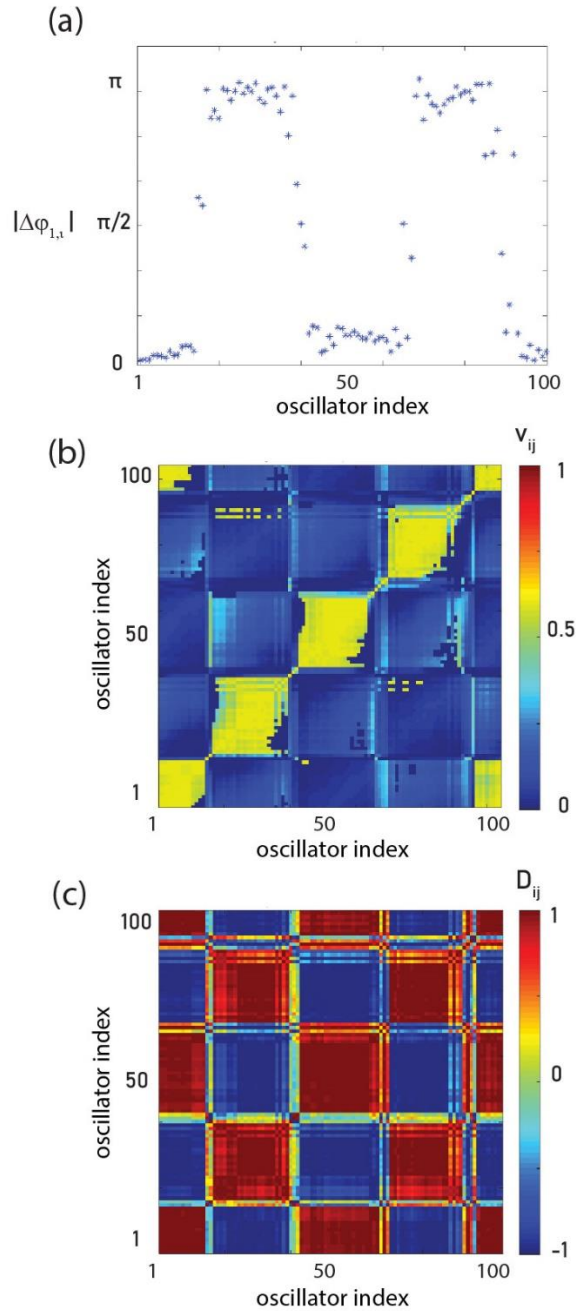

**Figure 11: Dissociation between structural and functional clusters for state{2,d} observed in scenario 2. Panel a)** phase offsets between every oscillator and the first ($|\Delta\varphi_{1,i}|$). Offsets reflect two anti-phase clusters. **Panel b)**, pairwise conduction velocity reflecting four structural clusters. **Panel c)**, pairwise functional connectivity reflecting two functional clusters.

This effect persists if both coupling strength and conduction velocity are dynamic. In this case, we observed that for a sufficiently large enhancement factor, which determines maximum conduction velocity, a single functional cluster exhibiting zero-mode synchronization emerges. However, structurally two clusters emerge with positive intra-cluster and negative inter-cluster connectivity (see figure 12). Adaptive conduction velocity thus allows for the functional integration of structurally segregated clusters.

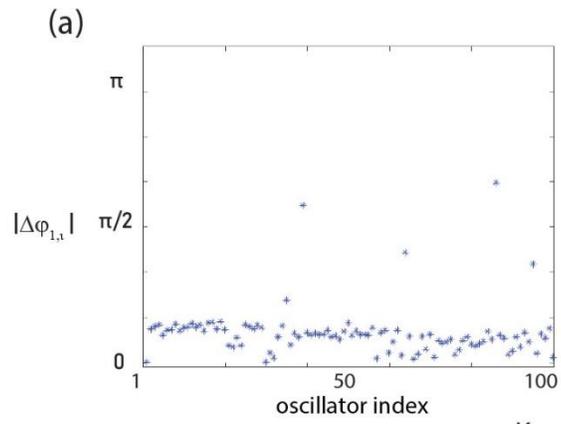

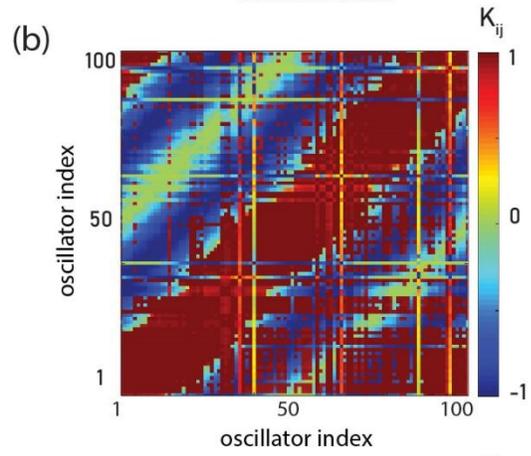

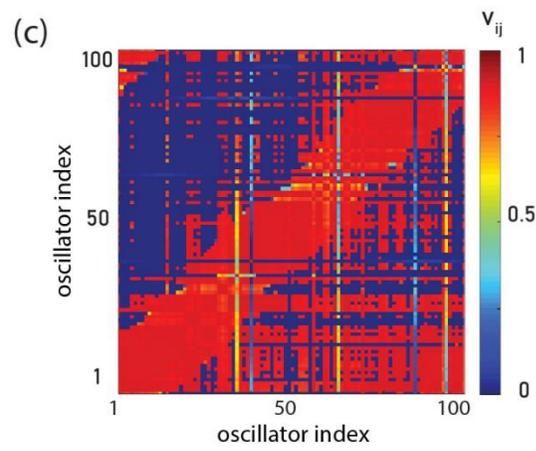

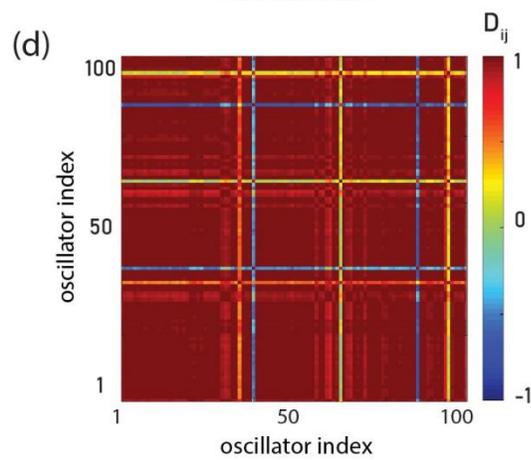

**Figure 12: Dissociation between structural and functional clusters for state{0,s} observed in scenario 3. Panel a)** phase offsets between every oscillator and the first ($|\Delta\varphi_{1,i}|$). Offsets reflect a single (global) cluster. **Panel b)**, pairwise structural connectivity reflecting two clusters. **Panel c)**, pairwise conduction velocity reflecting two clusters. **Panel d)**, pairwise functional connectivity reflecting a single cluster.

In light of neuroscientific evidence that myelination continues to exhibit adaptive changes even in the adult brain[28,41,42], our results highlight the importance of considering this factor in computational models of learning. For instance, our observation that dynamic conduction velocity provides the possibility for synchronization even in the context of fast learning highlights that adaptive myelination may have the capacity to compensate for synaptic effects that might otherwise desynchronize neural groups. Interestingly, the compensatory effect of dynamic conduction velocity could be observed in our simulations even when its rate of change is a factor of 10 slower than that of synaptic strength. This suggests that our findings are relevant for the biologically plausible scenario where myelin related changes lag behind changes in synaptic efficacy, as it may take up to several weeks of daily stimulation of neuronal axons before changes in myelination can be detected[43,44]. A role of slowly changing myelination in sharpening synchronization during neuronal communication would be in line with several theories in which rhythmic spike synchronization is thought to determine the efficiency of neural communication[45–48]. Our observation of a dissociation between structural and functional connectivity may thus be relevant for understanding whole-brain communication and help to further elucidate the relationship between brain structure and function.

Our results call for an investigation of the neuro-computational mechanisms allowing for activity- and experience-dependent modulations of adaptive myelination. Based on observations that white matter structural changes resemble synaptic changes to the extent that they depend on the frequency of neural co-activation[25–29,31,42,49], we implemented it as a Hebbian learning process. This is surely an over-simplification given that the control of myelination in adults, while incompletely understood, involves glia-neuronal interactions. We could not consider these here due to the simplicity of our model. Future work is thus needed to develop a more biologically appropriate learning mechanism and embed it in a model incorporating both types of cells. Nevertheless, our approach captures the most essential dynamical aspect of adaptive myelination, namely that conduction velocity of frequently used connections is strengthened while that of rarely used connections is weakened. Other simplifications of our work include the arrangement of oscillators along a circle and the lack of input. However, using these simplifications, we were able to decrease the complexity of computations and the number of parameters in order to plainly identify the influences of synaptic and myelin plasticity on collective behavior of oscillators. Future research will be necessary to investigate the contribution of realistic meso- and macroscopic network topology as well as of functionally relevant external stimulation.

## V. SUPPLEMENTARY MATERIAL

In the following, we present absolute phase offsets between every oscillator and the first, structural connectivity matrices (if coupling strengths are dynamic), functional connectivity matrices and conduction velocity matrices (if conduction velocities are dynamic) for every combination of learning parameters (i.e., $T$ and $\varepsilon_s$ for scenario I, and $\varepsilon_v$ and $\alpha_v$ for scenarios II and III) obtained from a single example simulation for each scenario. The parameter space of each example case is depicted in panel (a) of each figure.

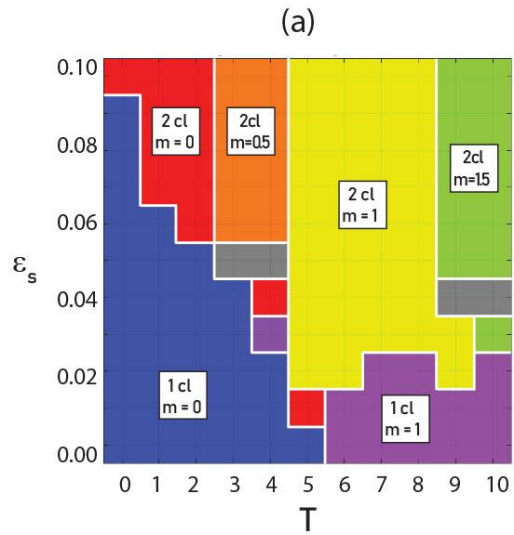

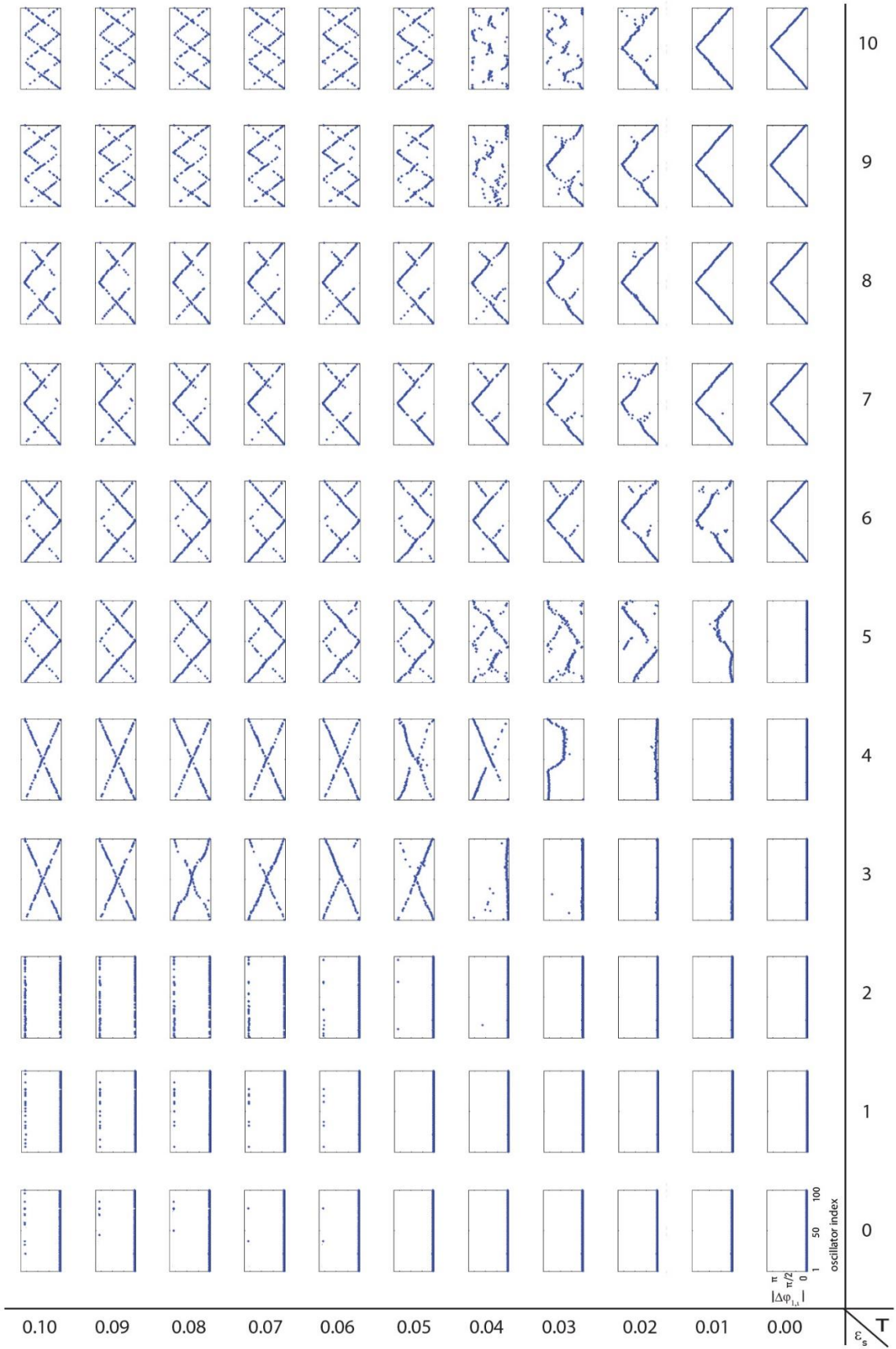

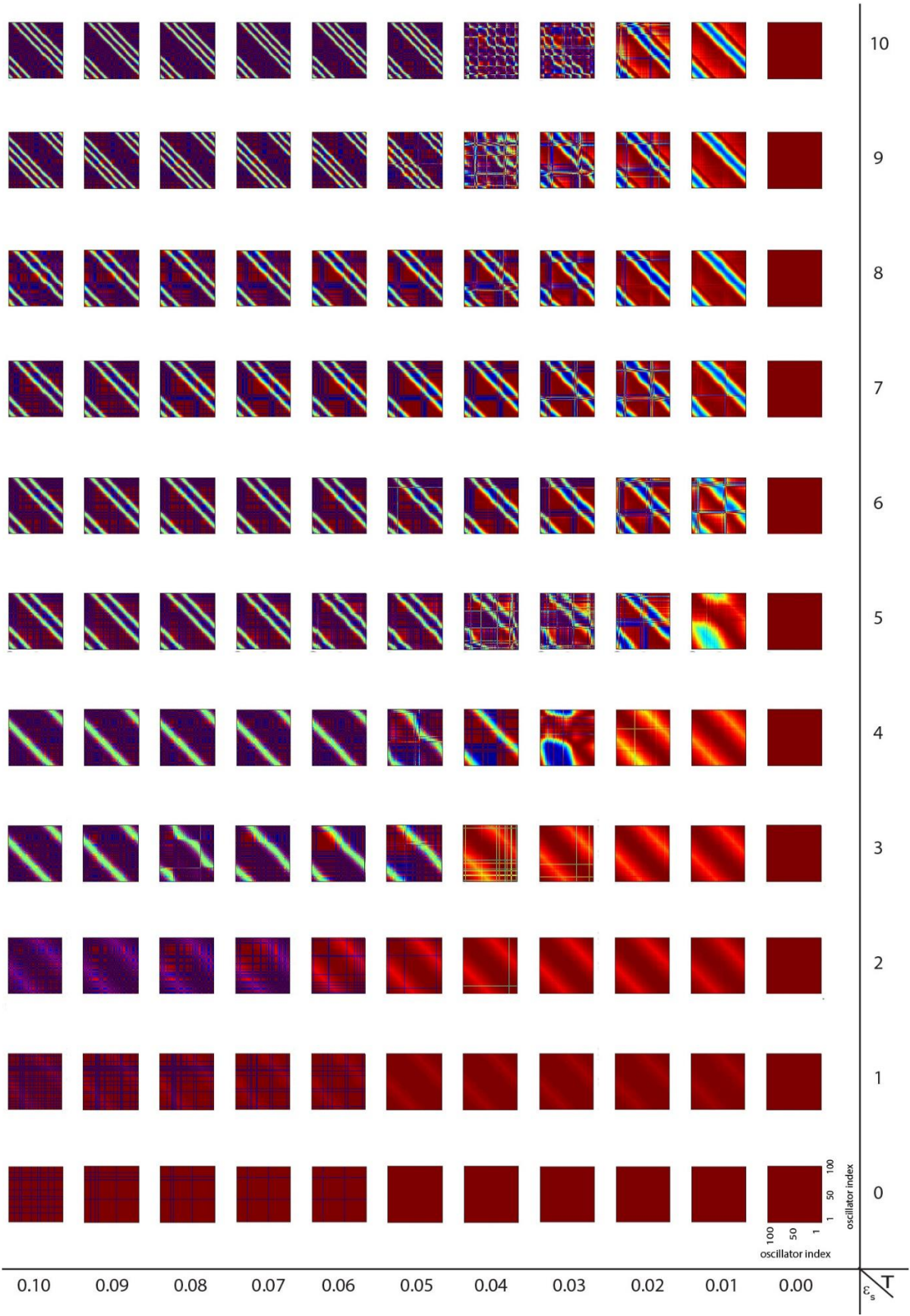

(d)

**Figure S1: Representation of structural and functional behavior characteristic of scenario I for every point of the parameter space. Panel a)** shows the color-coded state of coherent-wave mode of synchronization and cluster-formation obtained from a single simulation. **Panel b)** shows the absolute phase offsets between every oscillator and the first ($|\Delta\varphi_{1,i}|$) for all combinations of $\varepsilon_s$ and $T$. Phase offsets are averaged over the last 100 time steps of the simulation. **Panel c)** shows structural coupling matrices for all combinations of $\varepsilon_s$ and $T$. **Panel d)** shows functional connectivity matrices for all combinations of $\varepsilon_s$ and $T$. Matrix elements are averages over the last 100 time steps of the simulation in panels c and d. The learning enhancement factor $\alpha_s$ is fixed at 1 and all values of the connectivity matrix $K$ were initialized to $\alpha_s$.

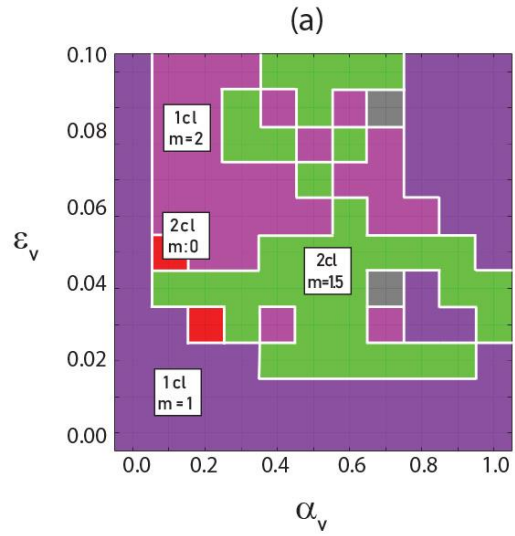

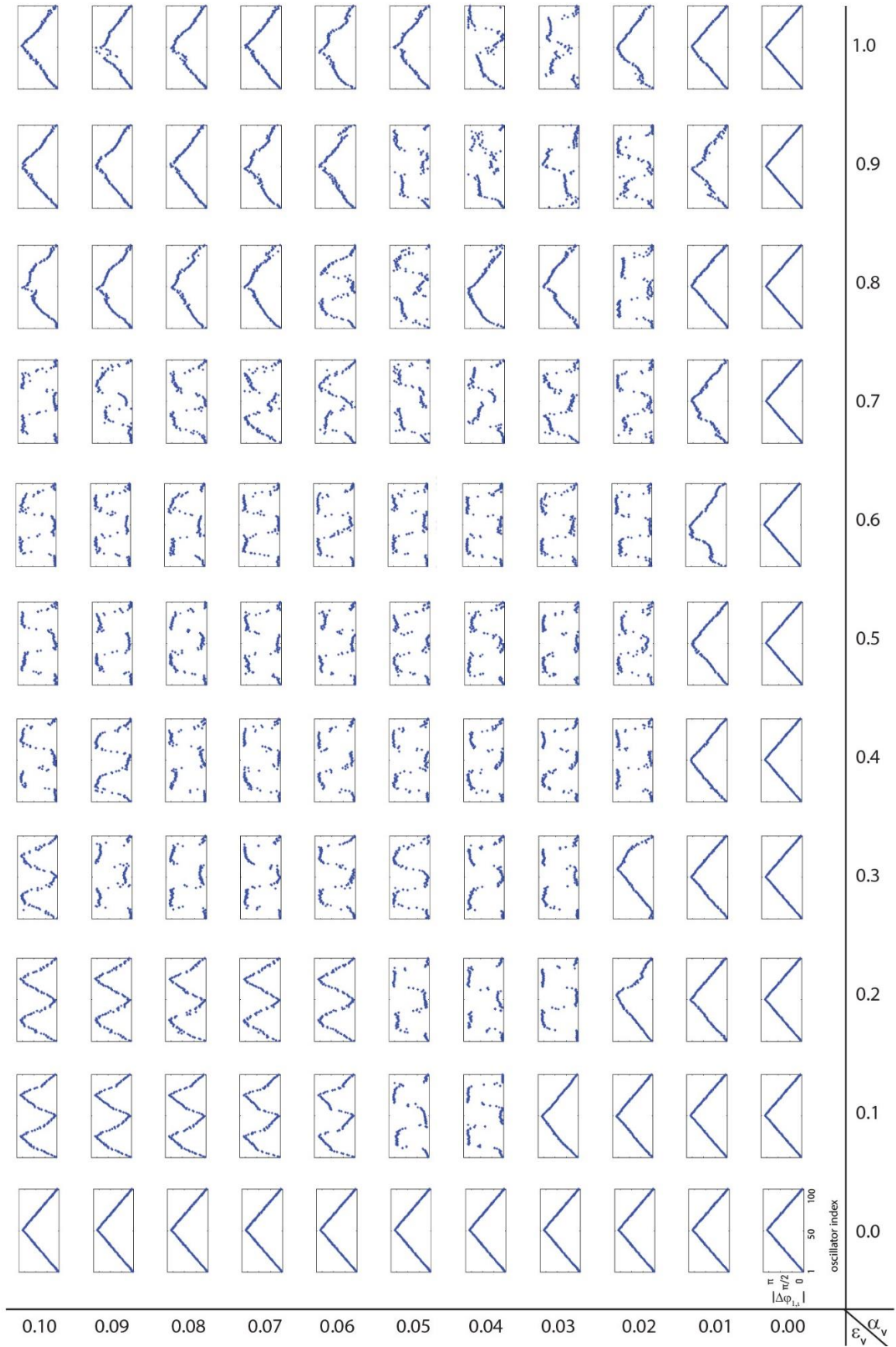

(c)

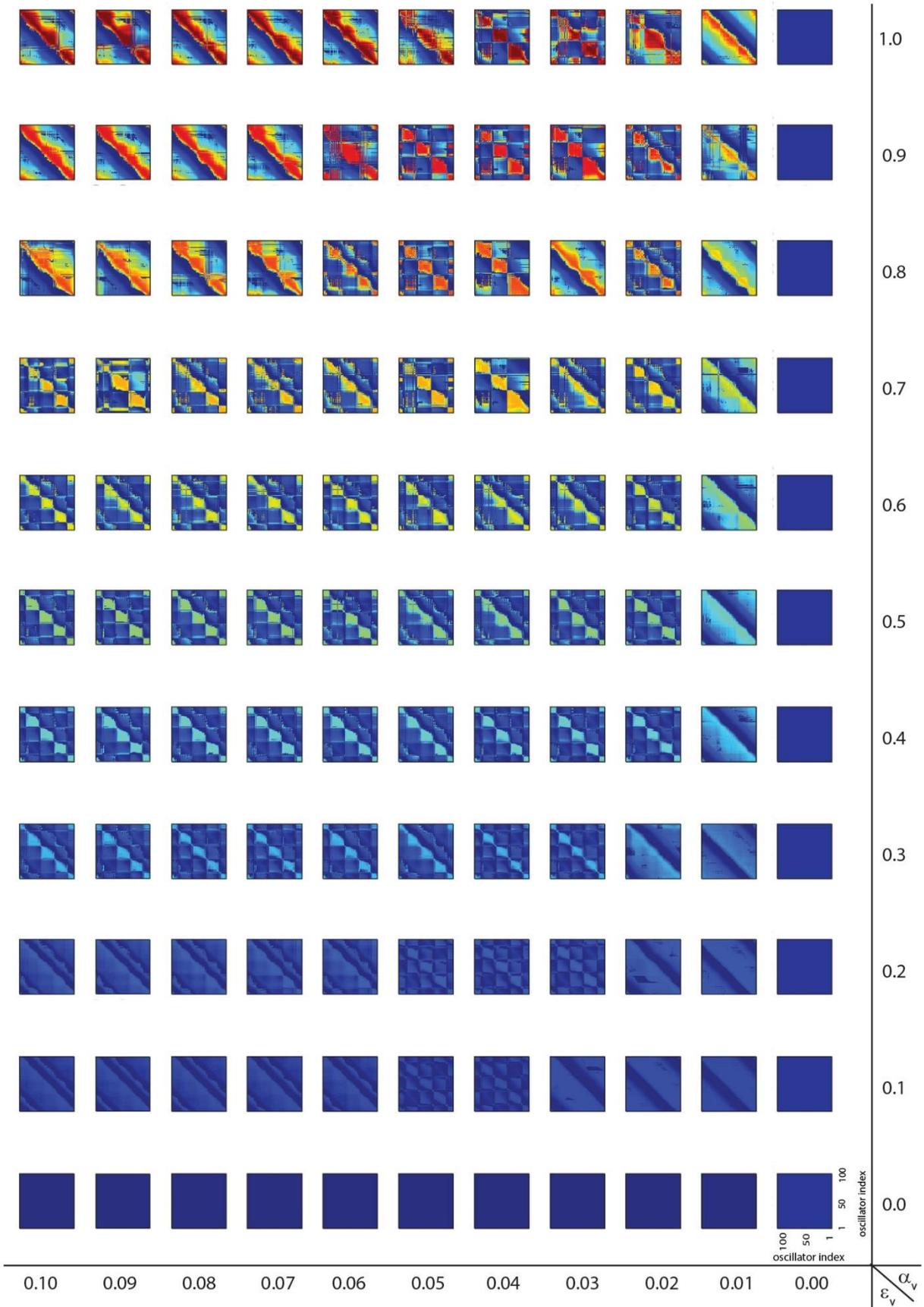

(d)

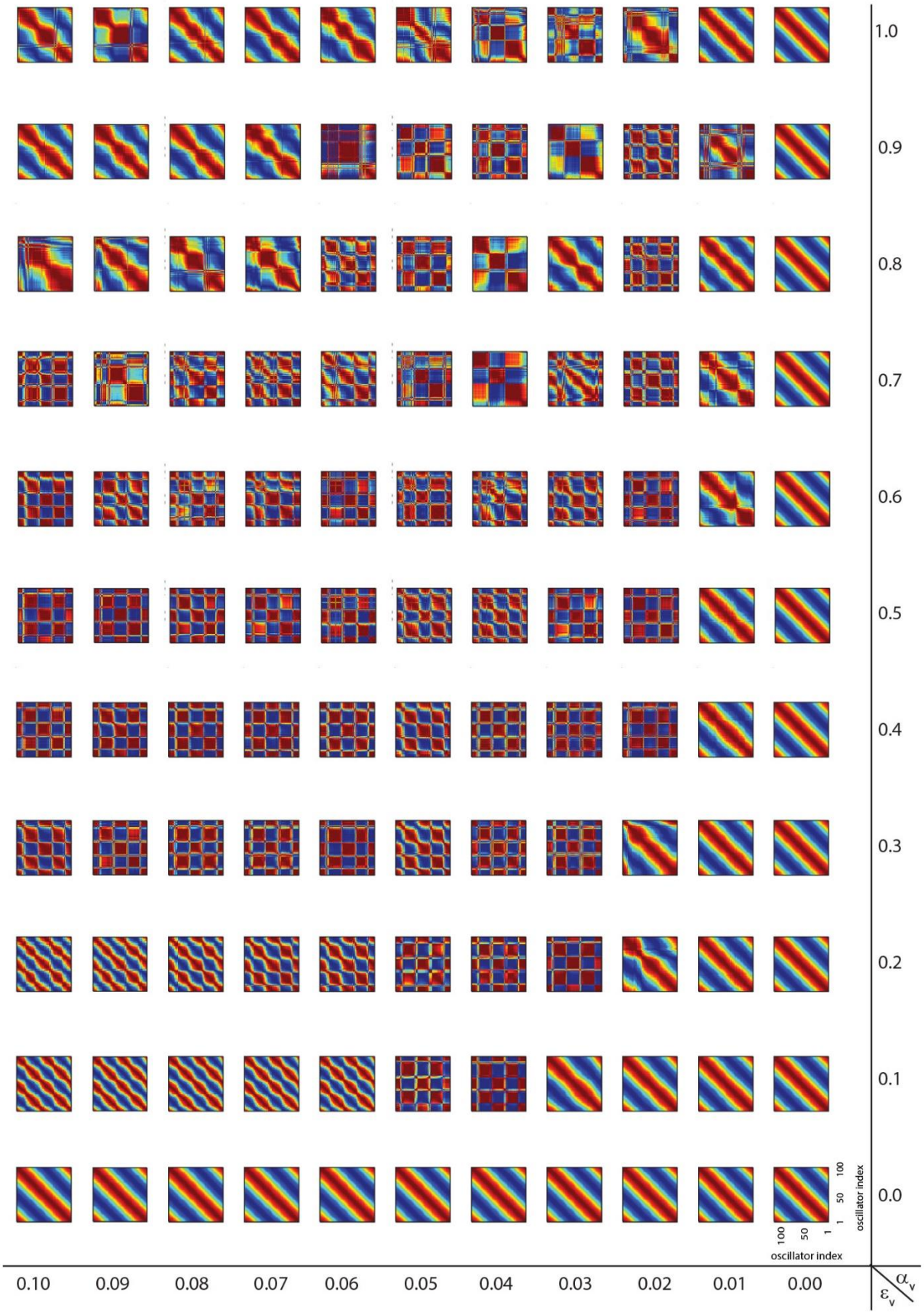

**Figure S2: Representation of structural and functional behavior characteristic of scenario II for every point of the parameter space. Panel a)** shows the color-coded state of coherent-wave mode of synchronization and cluster-formation obtained from a single simulation. **Panel b)** shows absolute phase offsets between every oscillator and the first ($|\Delta\varphi_{1,i}|$) for all combinations of $\varepsilon_v$ and $\alpha_v$. Phase offsets are averaged over the last 100 time steps of the simulation. **Panel c)** shows conduction velocity matrices for all combinations of $\varepsilon_v$ and $\alpha_v$. **Panel d)** shows functional connectivity matrices for all combinations of $\varepsilon_v$ and $\alpha_v$. Matrix elements are averages over the last 100 time steps of the simulation in panels c and d. The connections' learning rate $\varepsilon_s$ and the learning enhancement factor $\alpha_s$ are fixed at 0 and 1 respectively, and all values of the connectivity matrix K were fixed at $\alpha_s$. The pairwise conduction velocities $v_{ij}$ are initialized at 0.14, equivalent with the condition where $T \cong 7$.

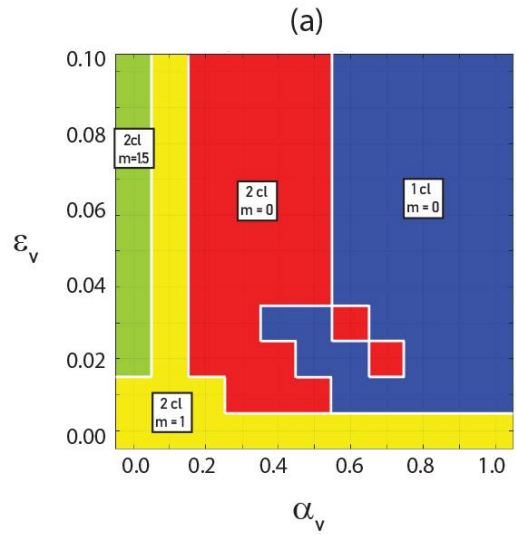

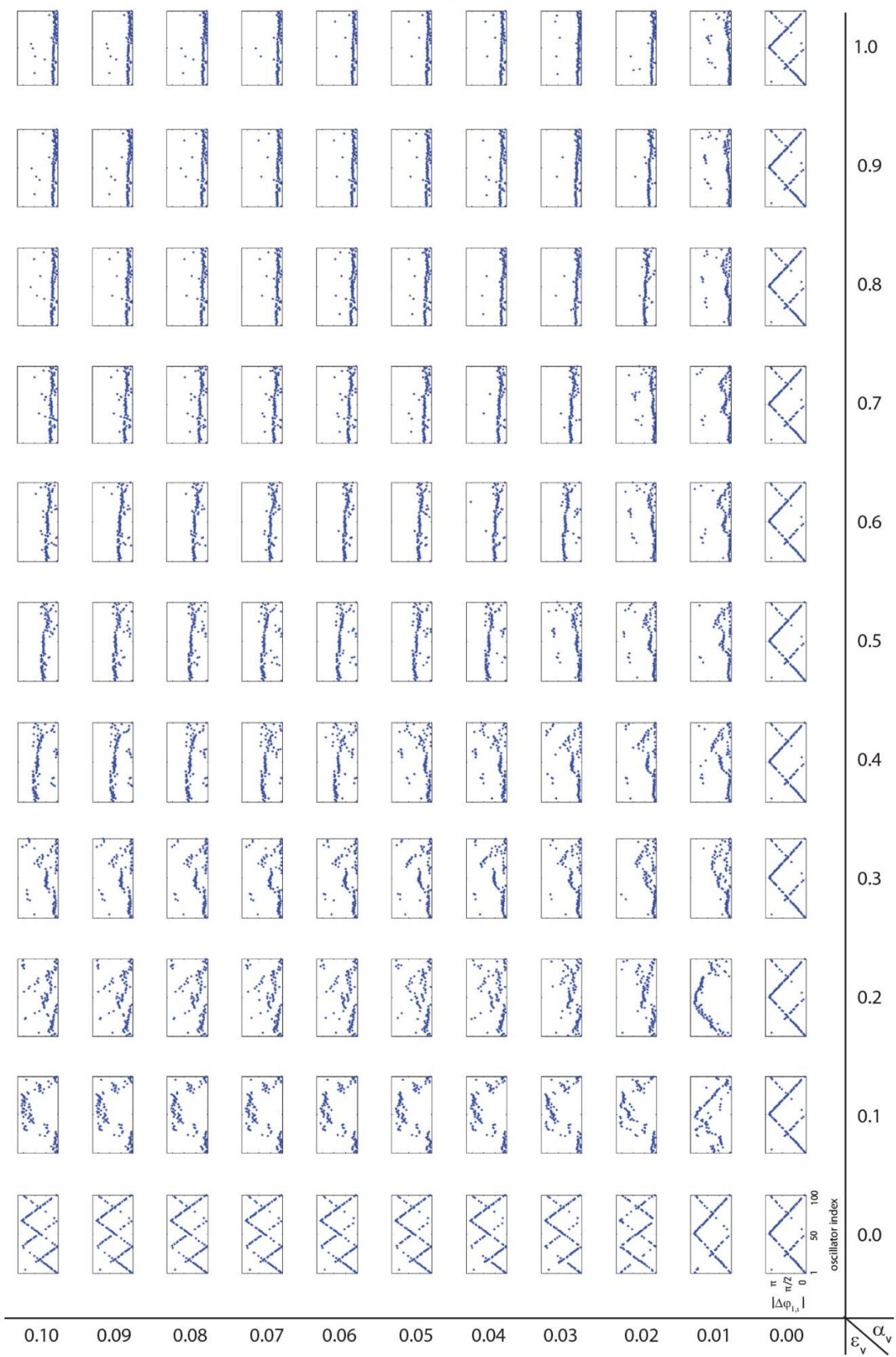

(c)

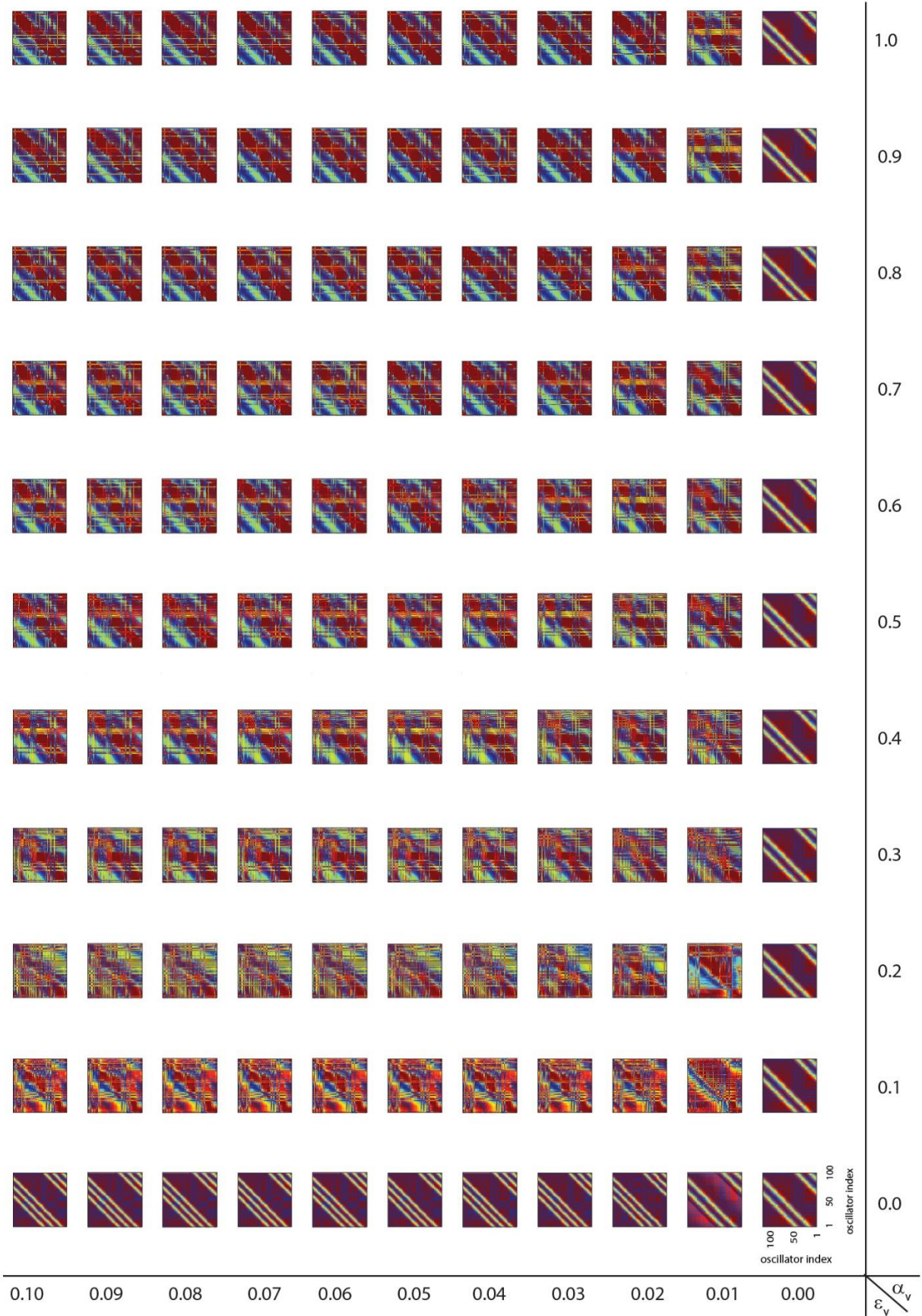

(d)

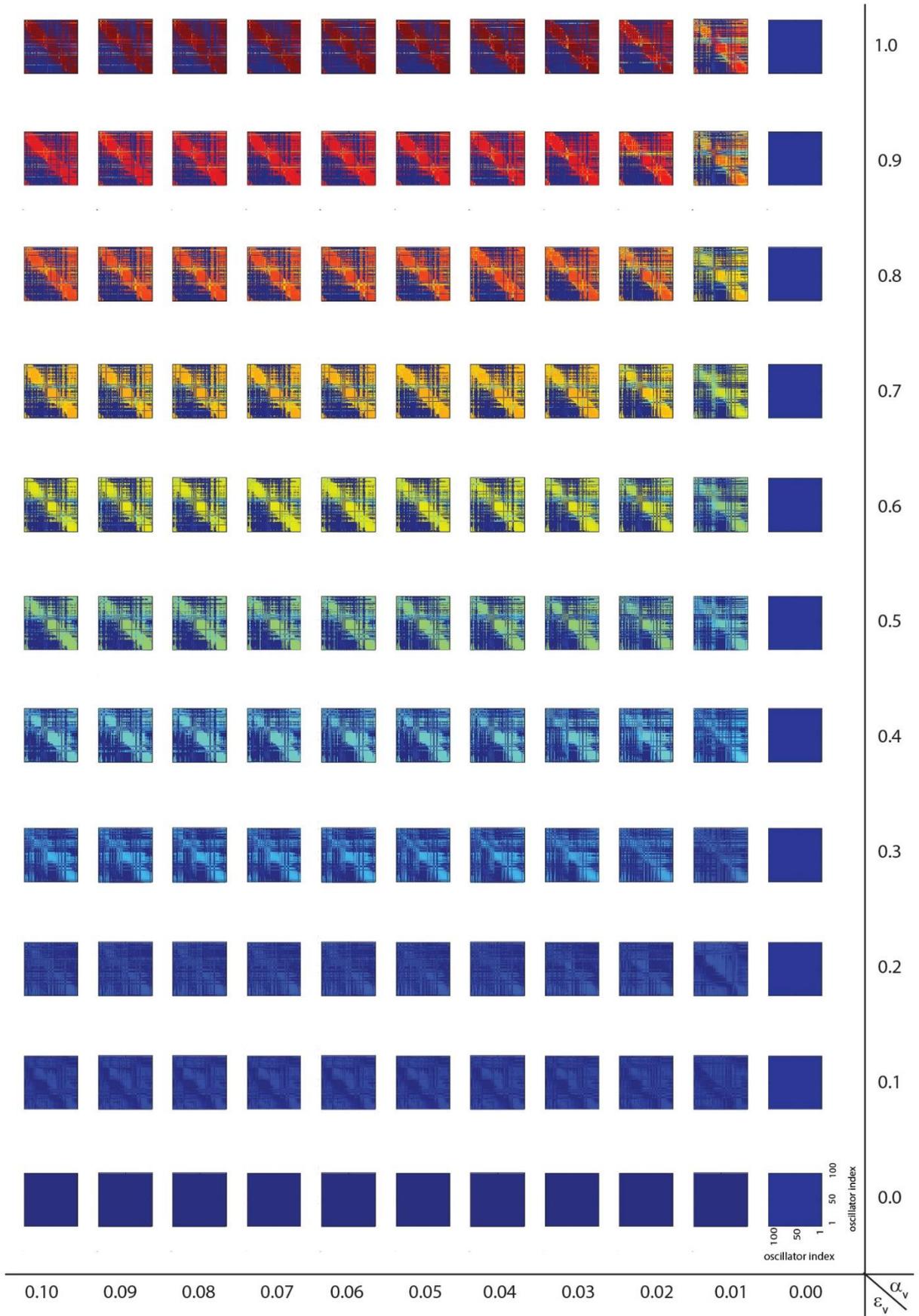

(e)

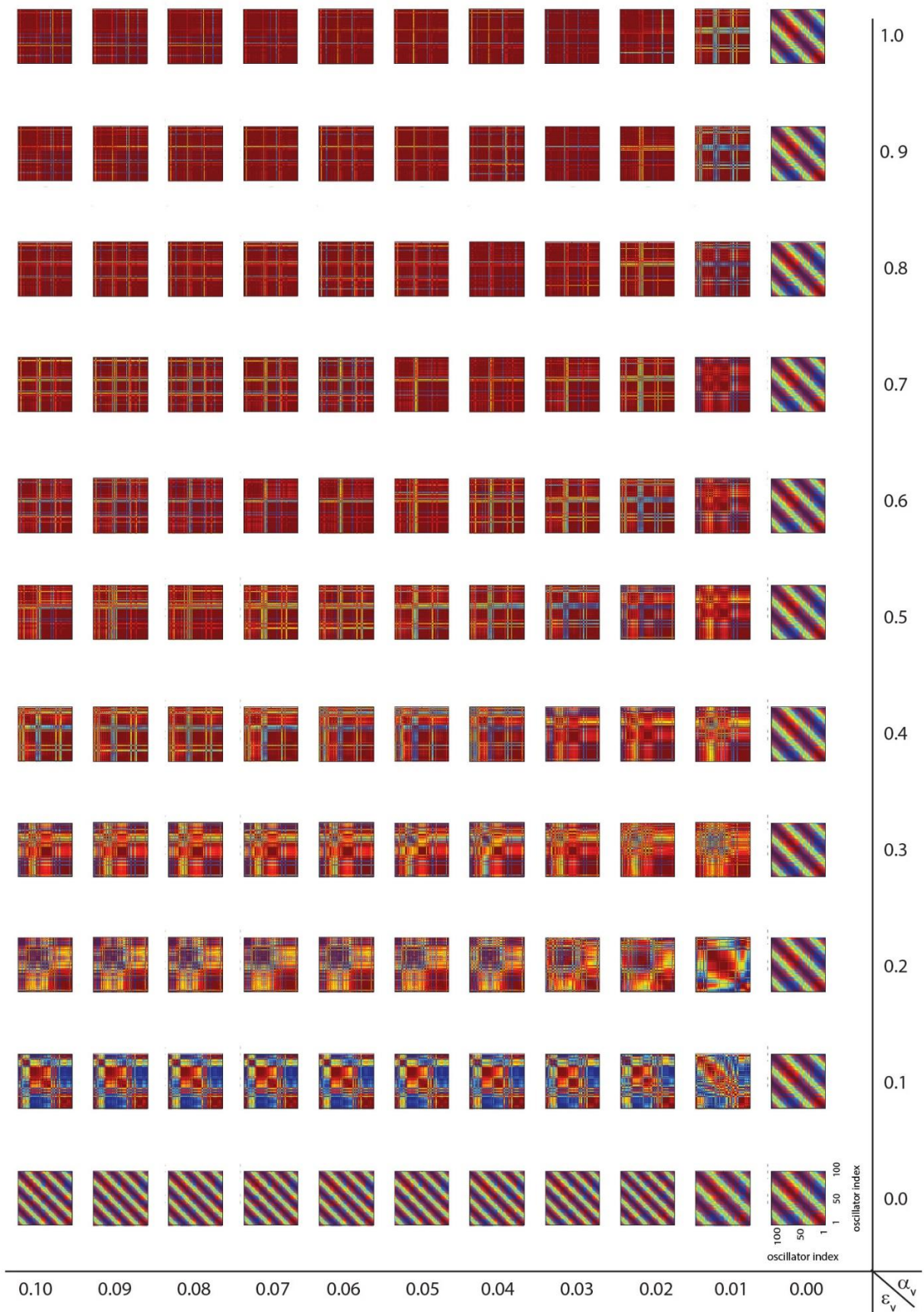

**Figure S 3: Representation of structural and functional behavior characteristic of scenario III for every point of the parameter space. Panel a)** shows the color-coded state of coherent-wave mode of synchronization and cluster-formation obtained from a single simulation. **Panel b)** shows absolute phase offsets between every oscillator and the first ($|\Delta\varphi_{1,i}|$) for all combinations of $\varepsilon_v$ and $\alpha_v$. Phase offsets are averaged over the last 100 time steps of the simulation. **Panel c)** shows structural coupling matrices for all combinations of $\varepsilon_v$ and $\alpha_v$. **Panel d)** shows conduction velocity matrices for all combinations of $\varepsilon_v$ and $\alpha_v$. **Panel e)** shows functional connectivity matrices for all combinations of $\varepsilon_v$ and $\alpha_v$. Matrice elements are averages over the last 100 time steps of the simulation in panels c, d and e. The connections' learning rate $\varepsilon_s$ and the learning enhancement factor $\alpha_s$ are fixed at 0.1 and 1 respectively, and all values of the connectivity matrix K were fixed at $\alpha_s$. The pairwise conduction velocities $v_{ij}$ are initialized at 0.14, equivalent with the condition where $T \cong 7$.

## VI. ACKNOWLEDGMENTS

Author PDW was supported by an NWO VICI grant (453.04.002). Author MS was funded by the European Union's Horizon 2020 Research and Innovation Program under Grant Agreement No. 737691 (HBP SGA2). Author MM was supported by an NWO VENI grant (451.15.012). This work was supported by the Dutch Province of Limburg.